\documentclass[pdftex,twocolumn,epjc3]{svjour3}          

\RequirePackage[T1]{fontenc}

\smartqed  

\RequirePackage{graphicx}
\RequirePackage{mathptmx}      
\RequirePackage{flushend}
\RequirePackage[numbers,sort&compress]{natbib}
\RequirePackage[colorlinks,citecolor=blue,urlcolor=blue,linkcolor=blue]{hyperref}

\usepackage{amssymb}
\usepackage{marvosym}
\usepackage{amsmath}
\usepackage{multirow}
\usepackage{fancyhdr}
\usepackage{rotate}

\usepackage{lineno} 

\renewcommand{\d}{\mathrm{d}}
\newcommand{\pt}{p_\mathrm{T}}
\newcommand{\mt}{m_\mathrm{T}}
\newcommand{\snn}{\sqrt{s_\mathrm{NN}}}

\journalname{Eur. Phys. J. C}

\begin{document}

\title{Nuclear dependence of light neutral meson production \\ in p-A collisions at 400~GeV with NA60}

\author{R.~Arnaldi\thanksref{addre1}\and
        K.~Banicz\thanksref{addre2,addre3}\and
        K.~Borer\thanksref{addre4}    	     
        J.~Castor\thanksref{addre5}\and         
        B.~Chaurand\thanksref{addre6}\and         
        W.~Chen\thanksref{addre7}\and         
        C.~Cical\`o\thanksref{addre8}\and         
        A.~Colla\thanksref{addre1,addre9}\and    	     
        P.~Cortese\thanksref{addre9,addre10}\and         
        S.~Damjanovic\thanksref{addre2,addre3}\and       
        A.~David\thanksref{addre2,addre11}\and       
        A.~De~Falco\thanksref{addre8,addre12}\and         
        A.~Devaux\thanksref{addre5}\and         
        L.~Ducroux\thanksref{addre13}\and        
        H.~En'yo\thanksref{addre14}\and   	     
        J.~Fargeix\thanksref{addre5}\and         
        A.~Ferretti\thanksref{addre1,addre9}\and         
        M.~Floris\thanksref{addre8,addre12}\and         
        A.~F\"orster\thanksref{addre2}\and         
        P.~Force\thanksref{addre5}\and    	     
        N.~Guettet\thanksref{addre2,addre5}\and       
        A.~Guichard\thanksref{addre13}\and        
        H.~Gulkanian\thanksref{addre15}\and        
        J.~M.~Heuser\thanksref{addre14}\and        
        P.~Jarron\thanksref{addre2}\and         
        M.~Keil\thanksref{addre2,addre11}\and  	     
        L.~Kluberg\thanksref{addre6}\and         
        Z.~Li\thanksref{addre7}\and         
        C.~Louren\c{c}o\thanksref{addre2}\and         
        J.~Lozano\thanksref{addre11}\and         
        F.~Manso\thanksref{addre5}\and         
        P.~Martins\thanksref{addre2,addre11}\and       
        A.~Masoni\thanksref{addre8}\and         
        A.~Neves\thanksref{addre11}\and    	     
        H.~Ohnishi\thanksref{addre14}\and        
        C.~Oppedisano\thanksref{addre1}\and         
        P.~Parracho\thanksref{addre2,addre11}\and       
        P.~Pillot\thanksref{addre13}\and        
        T.~Poghosyan\thanksref{addre15}\and        
        G.~Puddu\thanksref{addre12}\and    	     
        E.~Radermacher\thanksref{addre2}\and         
        P.~Ramalhete\thanksref{addre2,addre11}\and       
        P.~Rosinsky\thanksref{addre2}\and         
        E.~Scomparin\thanksref{addre1}\and         
        J.~Seixas\thanksref{addre11,addre16,addre17}\and         
        S.~Serci\thanksref{addre8,addre12}\and         
        R.~Shahoyan\thanksref{addre2,addre11}\and       
        P.~Sonderegger\thanksref{addre11}\and         
        H.~J.~Specht\thanksref{addre3}\and         
        R.~Tieulent\thanksref{addre13}\and        
        A.~Uras\thanksref{addre8,addre12,addre13}\thanksref{t1}\and 
        G.~Usai\thanksref{addre8,addre12}\thanksref{t1}\and    
        R.~Veenhof\thanksref{addre11}\and         
        H.~K.~W\"ohri\thanksref{addre2,addre11,addre12}\and
        (NA60 Collaboration)
}

\institute{INFN Torino, Italy\label{addre1}\and
           CERN, Geneva, Switzerland\label{addre2}\and
           Physikalisches Institut der Universit\"{a}t Heidelberg, Germany\label{addre3}\and
           University of Bern, Switzerland\label{addre4}\and
           LPC, Université Clermont Auvergne and CNRS-IN2P3, Clermont-Ferrand, France\label{addre5}\and
           LLR, Ecole Polytechnique and CNRS-IN2P3, Palaiseau, France\label{addre6}\and
           BNL, Upton, NY, USA\label{addre7}\and
           INFN Cagliari, Italy\label{addre8}\and
           Universit\`a di Torino, Italy\label{addre9}\and
           Universit\`a del Piemonte Orientale, Alessandria, Italy\label{addre10}\and
           Instituto Superior Tecnico, Dep. Fisica, Lisbon, Portugal\label{addre11}\and
           Universit\`a di Cagliari, Italy\label{addre12}\and
           IPN-Lyon, Univ.\ Claude Bernard Lyon-I and CNRS-IN2P3, Lyon, France\label{addre13}\and
           RIKEN, Wako, Saitama, Japan\label{addre14}\and
           YerPhI, Yerevan, Armenia\label{addre15}\and
           Center for Physics and Engineering of Advanced Materials (CeFEMA), Lisbon, Portugal\label{addre16}\and
           Laboratorio de Intrumenta\c{c}\~ao e Fisica experimental de Particulas (LIP), Lisbon, Portugal\label{addre17}
}

\thankstext[$\star$]{t1}{~Corresponding authors: antonio.uras@cern.ch (A.~Uras), \\ gianluca.usai@ca.infn.it (G.~Usai)}

\date{Received: date / Accepted: date}

\maketitle


\begin{abstract}
The NA60 experiment has studied low-mass muon pair production in
proton-nucleus collisions with a system of Be, Cu, In, W, Pb and~U
targets, using a 400~GeV proton beam at the CERN SPS. 
The transverse momentum spectra of the $\rho/\omega$ and $\phi$ mesons are measured in 
the full $\pt$ range accessible, from $\pt = 0$ up to 2~GeV/$c$. 
The nuclear dependence of the production cross sections
of the $\eta$, $\omega$ and $\phi$ mesons has been found to be consistent 
with the power law $\sigma_\mathrm{pA} \propto \mathrm{A}^\alpha$, 
with the $\alpha$ parameter increasing as a function of $\pt$ for all the particles, and 
an approximate hierarchy $\alpha_\eta \approx \alpha_\phi > \alpha_\omega$.
The cross section ratios $\sigma_\eta/\sigma_\omega$,
$\sigma_\rho/\sigma_\omega$ and $\sigma_\phi/\sigma_\omega$ have been studied as a function
of the size~A of the production target, and an increase of the
$\eta$ and $\phi$ yields relative to the $\omega$ is observed from p-Be
to p-U collisions. 
\end{abstract}

\section{Introduction}

\noindent The study of the production of low-mass vector and
pseudoscalar mesons in proton-nucleus (p-A) collisions represents a unique tool
to understand the role of cold nuclear matter in particle production mechanisms.
The nuclear dependence of both the transverse momentum spectra and the total production
cross sections is of special interest in this context, with the observations usually
interpreted in terms of phenomenological models, since a first-principle 
description based on non-perturbative QCD is not yet
available.

Proton-nucleus collisions have also become of special interest
in their own, as a tool to test the possible existence of 
in-medium modifications of the vector meson spectral functions in cold nuclear matter. 
The experimental evidence for such effects is at the
moment controversial, namely after the KEK-PS 
E325~\cite{Tabaru:2006az} and the CLAS~\cite{Nasseripour:2007mga,Wood:2010ei} experiments
reporting contradicting observations on possible $\rho$-meson broadening and mass-shift 
in cold nuclear matter.
In this context, a recent measurement in p-A collisions at 400~GeV by the NA60 experiment~\cite{Arnaldi:2016pzu}, exploiting 
the same data set considered in the present analysis, ruled out the existence of any significant 
cold-nuclear matter effect at the SPS energies on the line shapes of the light vector mesons 
$\rho$, $\omega$ and $\phi$, giving in particular, 
for the first time, a precise characterisation of the $\rho$-meson line shape in p-A collisions. 
The aforementioned results from NA60 represent the basis for the new results presented in this letter.

Measurements in p-A collisions also provide
a useful cold-nuclear-matter reference for the observations in heavy-ion collisions,
allowing for the study of particle production as a function of the size~A of the
nucleus.

Despite the points of interest mentioned above, there is a
general lack of high-precision and high-statistics measurements of low-mass 
vector and pseudoscalar mesons in proton-nucleus collisions at the
SPS energies, in particular for what concerns dilepton data. The
nuclear dependence of the production cross sections for $\pi^0$,
$\eta$ and $\omega$ mesons was investigated at the energy of
$\snn=29.1$~GeV by the CERES experiment~\cite{Agakishiev:1998mw}
in p-Be and p-Au collisions, and by the HELIOS
experiment~\cite{Veenhof:1993xt} in p-Be collisions. Results from
these experiments were compared to, or complemented the measurements
in proton-proton collisions, in particular those performed by the NA27
experiment~\cite{AguilarBenitez:1991yy} relative to the production of
$\pi^0$, $\eta$, $\omega$ and $\rho$ mesons at $\snn=27.5$~GeV.
On the other hand, there are no accurate measurements on the nuclear
dependence of $\phi$ production in proton-nucleus collisions at the
SPS. At somehow higher energy, $\bar{K}^{*0}$ and $\phi$ mesons have been
recently measured in p-C, p-Ti and p-W interactions at
$\snn=41.6$~GeV by HERA-B~\cite{Abt:2006wt}.\\

\noindent The NA60 experiment complemented its main In-In programme with a high-luminosity 
proton-nucleus run, exposing to a 400~GeV proton beam six target materials: Be, Cu, In, W,
Pb and~U. This allowed for a comprehensive and detailed study of the nuclear dependence of the
$\eta$, $\rho$, $\omega$ and $\phi$ meson production. The present letter is organised as follows. 
First, the $\pt$ spectra are presented for the $\rho/\omega$ and $\phi$ mesons, for which
the measurement could be performed down to zero~$\pt$ --- while 
the acceptance coverage limited the measurement of $\eta$-meson production 
(via the $\eta\to\mu\mu\gamma$ Dalitz decay channel) to $\pt > 0.6$~GeV/$c$.
The nuclear dependence of the production
cross sections of $\eta$, $\omega$ and $\phi$ mesons,
integrated over the $\pt$ regions available for each particle,
is then investigated in terms of the power
law $\sigma_\mathrm{pA} \propto \mathrm{A}^\alpha$. The
available statistics also allowed for a dedicated study of the $\alpha$ parameters as a
function of~$\pt$. Finally, particle production is further discussed via the study 
of the cross section ratios in the full phase space, with strangeness enhancement being
specifically addressed through the measurement of the nuclear dependence of the cross 
section ratios $\sigma_\phi/\sigma_\omega$ and $\sigma_\eta/\sigma_\omega$.

\section{Apparatus and event selection}
\label{sec:na60-experiment}

\noindent During the 2004 run, the NA60 experiment collected data with
a system of nine sub-targets of different nuclear spe\-ci\-es --- Be,
Cu, In, W, Pb and~U --- simultaneously exposed to an incident
400~GeV proton beam. The individual target thickness was 
chosen to have approximately a similar statistical
sample from each target. The total interaction length was 7.5\%. The
beam intensity during the run was 4-$5\times10^8$ protons per second.

A general description of the NA60 apparatus can be found for
example in~\cite{Arnaldi:2008er}, while in~\cite{Arnaldi:2016pzu} some additional, 
specific details can be found, 
relevant to the setup used during the proton run of interest for the present analysis.

The produced dimuons are identified and measured by the muon spectrometer,
composed of a set of Multi-Wire Proportional Chamber (MWPC) tracking stations, trigger scintillator
hodoscopes, a toroidal magnet and a hadron absorber. 
The material which stops the hadrons also induces multiple scattering and
energy loss on the muons, degrading the mass resolution of the
measurement made in the spectrometer. To overcome this problem, NA60
already measures the muons before the absorber with a vertex spectrometer, 
made of pixel silicon detectors. 
The muon tracks reconstructed in the muon spectrometer are
extrapolated back to the target region and matched to the
tracks reconstructed in the vertex spectrometer. This is done
comparing both their angles and momenta, requiring a {\it matching}
$\chi^2$ less than 3. Once identified, the muons are refitted using
the joint information of the muon and the vertex spectrometers.  These
tracks will be referred to as \emph{matched muons}.  Muon pairs of
opposite charge are then selected.  The matching technique improves
significantly the signal-to-background ratio and the dimuon mass
resolution.  Because of the heavier absorber setup, the mass
resolution is slightly worse than during the In-In run: 30-35 MeV/$c^2$
(depending on target position) at the $\omega$ mass (against 23
MeV/$c^2$ during the In-In run), and the $\pt$ coverage is reduced 
towards the dimuon mass threshold.
\subsection{Target identification}
\noindent In order to study the nuclear dependence of the yields and
the kinematics for the particles produced in the collisions, the
identification of the production target is mandatory.  In principle,
one could consider the origin of the matched dimuon alone, defined as
the point of closest approach of its two muon tracks. However, this origin
has a limited spatial resolution; in particular, the error associated
to the $z$-position of the dimuon's origin becomes comparable or even
larger than the typical semi-distance between targets (5~mm) for
masses below 0.45 GeV/$c^2$. In order to overcome these difficulties,
the vertices reconstructed by using all the tracks measured in the
silicon telescope (VT vertices) are considered, requiring that the two muons are
attached to the same VT vertex, or to two different VT vertices
falling within the same target. 
The loss of statistics, as studied by Monte Carlo simulations, varies
from $\sim40\%$ for the $\phi$ mass region, to $\sim70\%$ in the dimuon mass 
region below 0.45 GeV/$c^2$ dominated by the Dalitz decay of the $\eta$.  
After applying this selection, $\sim80\,000$ muon pairs are
left. The algorithm fails to associate the correct target in
$\sim2\%$ of events at the $\phi$ mass, $\sim5\%$ at the $\omega$ mass
and $\sim15\%$ in the $\eta$ Dalitz region: this systematic
effect has been studied by means of the same Monte Carlo simulations, and
corrected for in the final results.

\subsection{Background treatment}
\noindent The small amount of combinatorial background
(originating from $\pi$ and $K$ decays) is subtracted from the raw dimuon sample: 
its shape is estimated with an event-mixing technique, while its
normalization is established fixing the like-sign (LS) component
coming out from the mixing to the LS component of the data (containing
no signal from correlated pairs at the SPS energies). In the considered proton-nucleus data,
the background accounts for less than $10\%$ of the integrated mass spectrum below
1.4~GeV$/c^2$. The comparison between mixed and real sample, in turn, gives an
average uncertainty of $10\%$ at most, for both the $(++)$ and the
$(--)$ components; because of the absolute low level of the
background and because of its smooth mass profile, this uncertainty hardly
affects the extraction of the signal from the considered dimuon sources.

The background from fake track matches, which could arise at high multiplicities
from the association of a muon track  to more than
one track in the vertex telescope with an acceptable matching
$\chi^2$, is significantly lower than the combinatorial background. For this
reason, its contribution is almost negligible in the proton-nucleus data --- being in any case
taken automatically into account by the overlay Monte Carlo technique adopted for the simulations.

\section{Monte Carlo simulations, acceptance and reconstruction efficiency}

\noindent The electromagnetic decays of the light, neutral pseudoscalar and
vector mesons ($\eta$, $\eta'$, $\rho$, $\omega$ and $\phi$) are the
dominating processes at the lower end of the dimuon mass spectrum (from the threshold
to the $\phi$ mass region),
adding to the continuum spectrum via their Dalitz decays and/or
giving rise to distinct peaks via their 2-body decays. This hadronic
decay cocktail was simulated with the NA60 Monte Carlo generator
Genesis~\cite{genesis}. The input parameters for the kinematic
distributions of the generated processes have been tuned by comparison
with the real data, by means of an iterative procedure ensuring
self-consistency to the analysis. 

The transverse momentum spectra 
used in the simulations are taken from the analysis itself.
The rapidity distributions in the center of mass frame were
generated according to the expression $\d N/\d y \propto 1/\cosh^2
(ay)$, similar to a Gaussian of width $\sigma =
0.75/a$, where $a$ describes the empirical functional mass dependence of the width with 
values of about 0.5 and 0.75 at the masses of 0.14~GeV/$c^2$ ($\pi^0$)
and 1~GeV/$c^2$, respectively~\cite{genesis}.  This simple
parameterisation has been used by several experiments, since it
describes reasonably well existing measurements~\cite{Afanasev:2000uu,Alber:1997sn}.  

The muon angular distributions also entering the simulations are assumed
to be isotropic for the 2-body decays, while 
the angular anisotropies of the Dalitz decays, expected to be the same for
the pseudo-scalar ($\eta,~\eta'$) and vector ($\omega$) mesons~\cite{Bratkovskaya:1996nf}, 
are described by the distribution:
\begin{equation}
	f(\theta) = 1 + \cos^2\theta + \left( \frac{2m_\mu}{M} \right)^2 \sin^2\theta~,
\end{equation}
where $M$ is the mass of the virtual photon, $m_\mu$ the mass of the muon,
and $\theta$ the angle between the positive muon and the momentum of the parent meson
in the rest-frame of the virtual photon~\cite{Anastasi:2016hdx}. As was explicitly verified, 
the dimuon acceptance for the Dalitz decays is practically unaffected by the
character of the angular distribution, due to the fact that the anisotropy 
is strongly smeared out in the laboratory frame.

For the mass line shapes of the narrow resonances $\eta$,
$\omega$ and $\phi$, a modified relativistic Breit-Wigner
parametrization was used, first proposed by G.J.~Gounaris and
J.J.~Sakurai~\cite{Gounaris:1968mw}, with widths and masses
taken from the Particle Data Group~(PDG) tables~\cite{Nakamura:2010zzi}. 
For the broad $\rho$ meson the following parameterisation was used~\cite{Knoll:1998iu}:

\begin{footnotesize}
\begin{equation}
   \frac{\d N}{\d M} \propto \frac{\sqrt{1-\frac{4m^2_\mu}{M^2}}\left( 1+\frac{2m^2_\mu}{M^2}\right)\left( 1-\frac{4m^2_\pi}{M^2}\right) ^{3/2}}
{\left( m^2_{\rho}-M^2\right)^2+m^2_\rho \mathrm{\Gamma}^2_\rho(M)} \left(MT\right)^{3/2} e^{-\frac{M}{T_\rho}}
\end{equation}
\end{footnotesize}

\noindent with a mass dependent width 

\begin{footnotesize}
\begin{equation}
\mathrm{\Gamma}_\rho(M)=\mathrm{\Gamma}_{0\rho} \frac{m_\rho}{M}\left(
\frac{M^2/4-m^2_\mu} {m^2_\rho/4-m^2_\mu}\right)^{3/2}=\mathrm{\Gamma}_{0\rho}
\frac{m_\rho}{M} \left( \frac{q} {q_0}\right)^{3}.
\end{equation}
\end{footnotesize}
\noindent The parameters of the $\rho$ parameterisation were either 
fixed to the PDG values~\cite{Nakamura:2010zzi}, or extracted from the 
data themselves, as extensively discussed in~\cite{Arnaldi:2016pzu}.

The dimuon mass distributions of the $\eta$ and $\omega$ Dalitz decays
are described by the QED expectations for point-like particles~\cite{Kroll:1955zu},
corrected by the form factors also extracted from the data themselves~\cite{Arnaldi:2016pzu}.

The semimuonic simultaneous decays from  $D\bar D$ mesons produce
a smooth open charm continuum with a maximum at around 1~GeV/$c^2$. They
were simulated with PYTHIA~6.4~\cite{Sjostrand:2006za}, with the mass of charm quark 
set to $m_c = 1.5$~GeV/$c^2$ and the primordial momentum of the interacting partons 
generated according to a Gaussian distribution of variance $k_\mathrm{T}^2 = 1.0$~(GeV/$c$)$^2$.\\

\noindent The Monte Carlo simulations were performed using the overlay
technique, which consists of superimposing a Monte-Carlo-generated muon
pair onto real events, in order to realistically simulate the underlying hadronic
event together with the detector specific behaviour. A real event is
read, chosen among the reconstructed data collected by the experiment,
containing a high-mass matched dimuon (within the $J/\psi$ mass
window, the two stiff muons guaranteeing an unambiguous identification of the interaction target) 
whose vertex is imposed to be the origin of the generated muon pair. 
Alternatively, dimuons whose vertex has the $z$-coordinate determined with an uncertainty
smaller than 3~mm were also used. This second choice, applying weaker
conditions on the vertex candidates, has been considered for
systematic checks in the analysis.  The muon pair produced in the
simulation is tracked through the NA60 apparatus, using
GEANT3~\cite{Brun:1987ma}. Starting from the ensemble of simulated and real hits, the events
in which a muon pair gave rise to a trigger were reconstructed using
the same reconstruction settings used for the real data.  To make the
MC simulation as realistic as possible, the MC tracks leave a signal
in a given pixel plane with a probability proportional to the plane
efficiency as estimated from the analysis of the real data.\\

\noindent The $\pt$ dependence of the dimuon acceptance results from a complex convolution
of several factors: the geometric acceptance of the vertex telescope,
the energy loss of muons in the hadron absorber, the geometric requirements
embedded in the trigger logic of the muon spectrometer (which requires
two muons to belong to different sextants), the combination of the
magnetic fields provided by the dipole magnet in the vertex region and
the toroidal magnet in the muon spectrometer after the absorber. 
An additional loss of factor up to $\sim 2$ is caused by the reconstruction efficiency.
\begin{figure*}[tbh] 
    \begin{center}
    \includegraphics[width=0.32\textwidth]{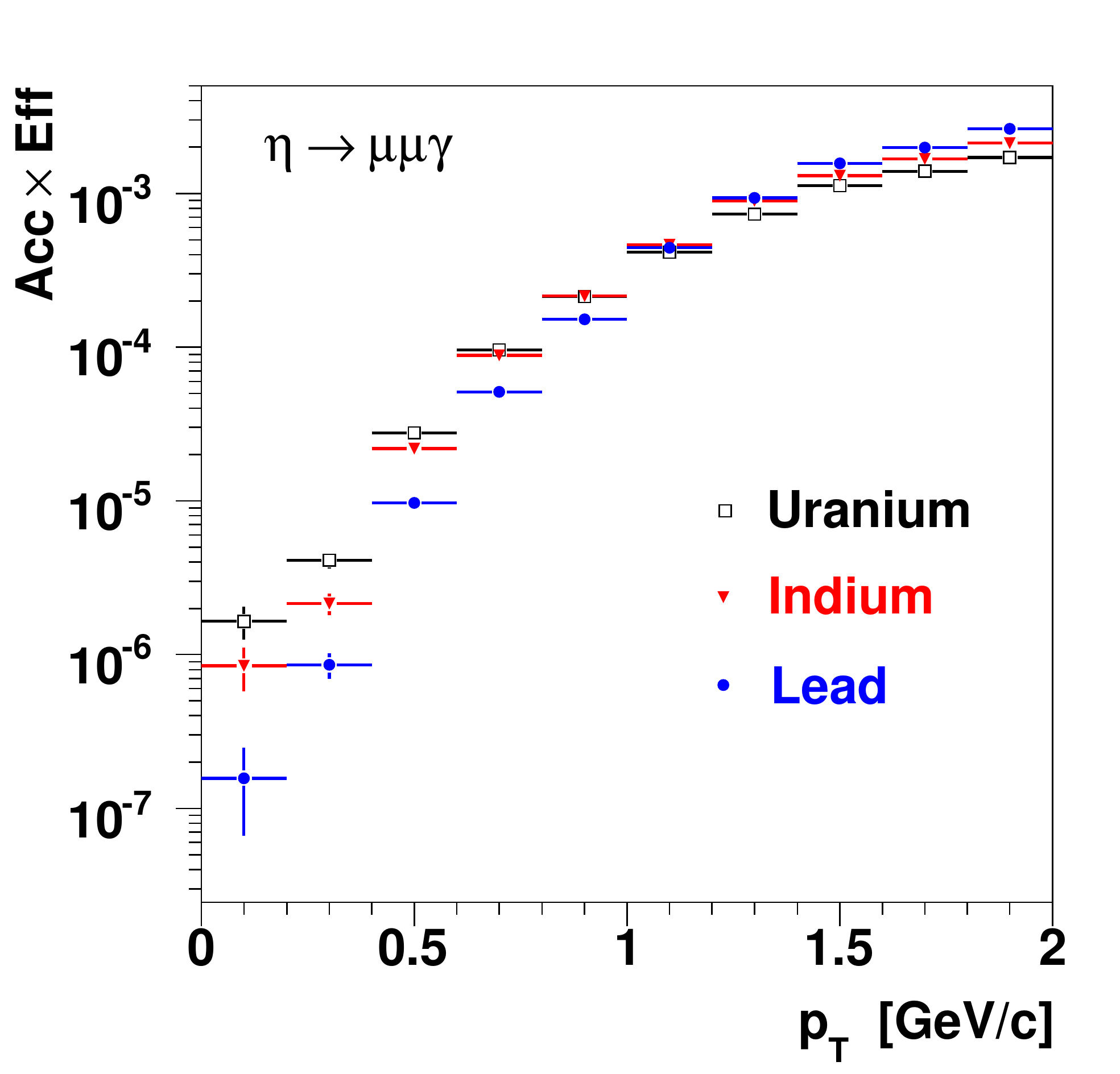} 
    \includegraphics[width=0.32\textwidth]{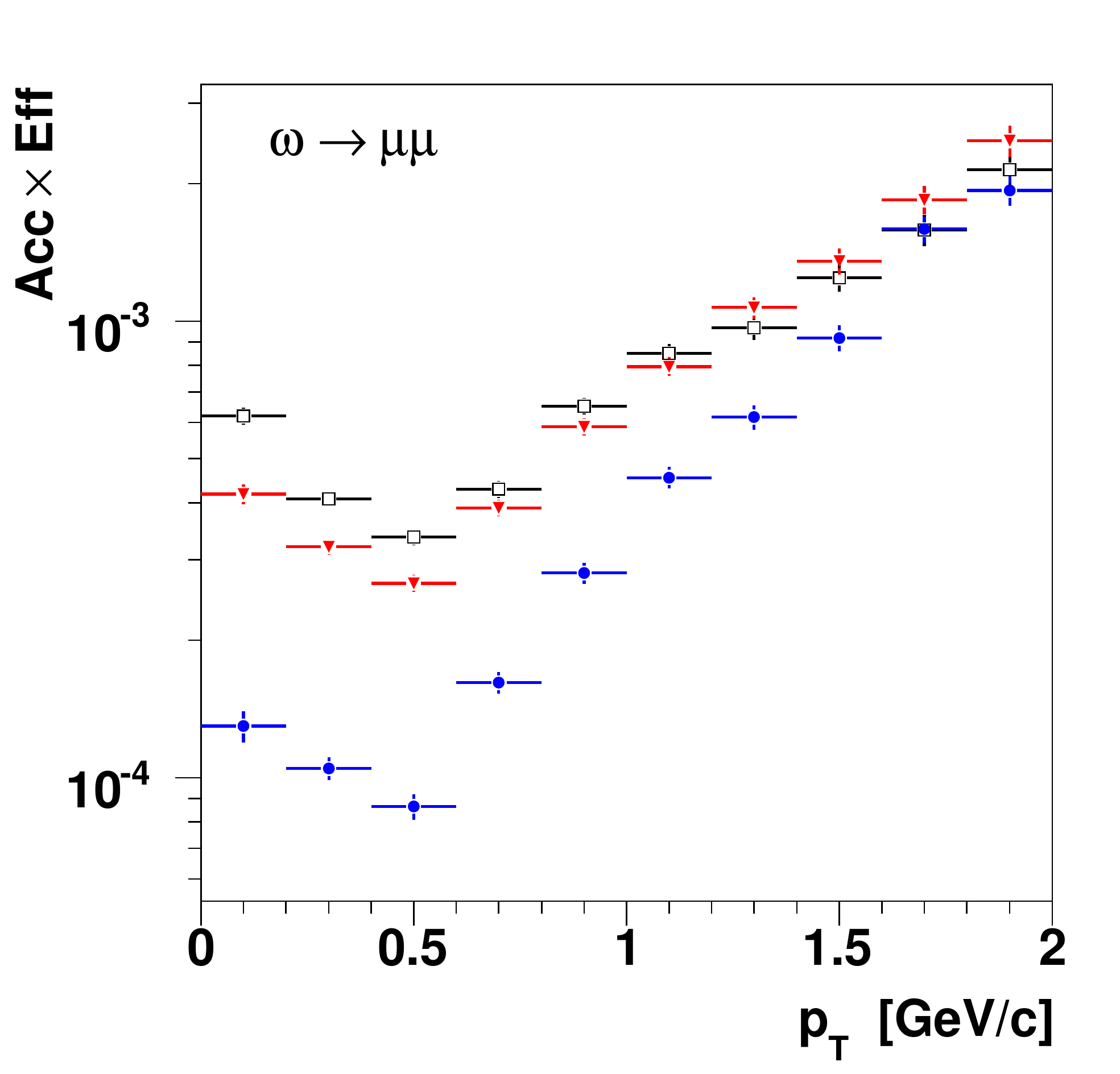} 
    \includegraphics[width=0.32\textwidth]{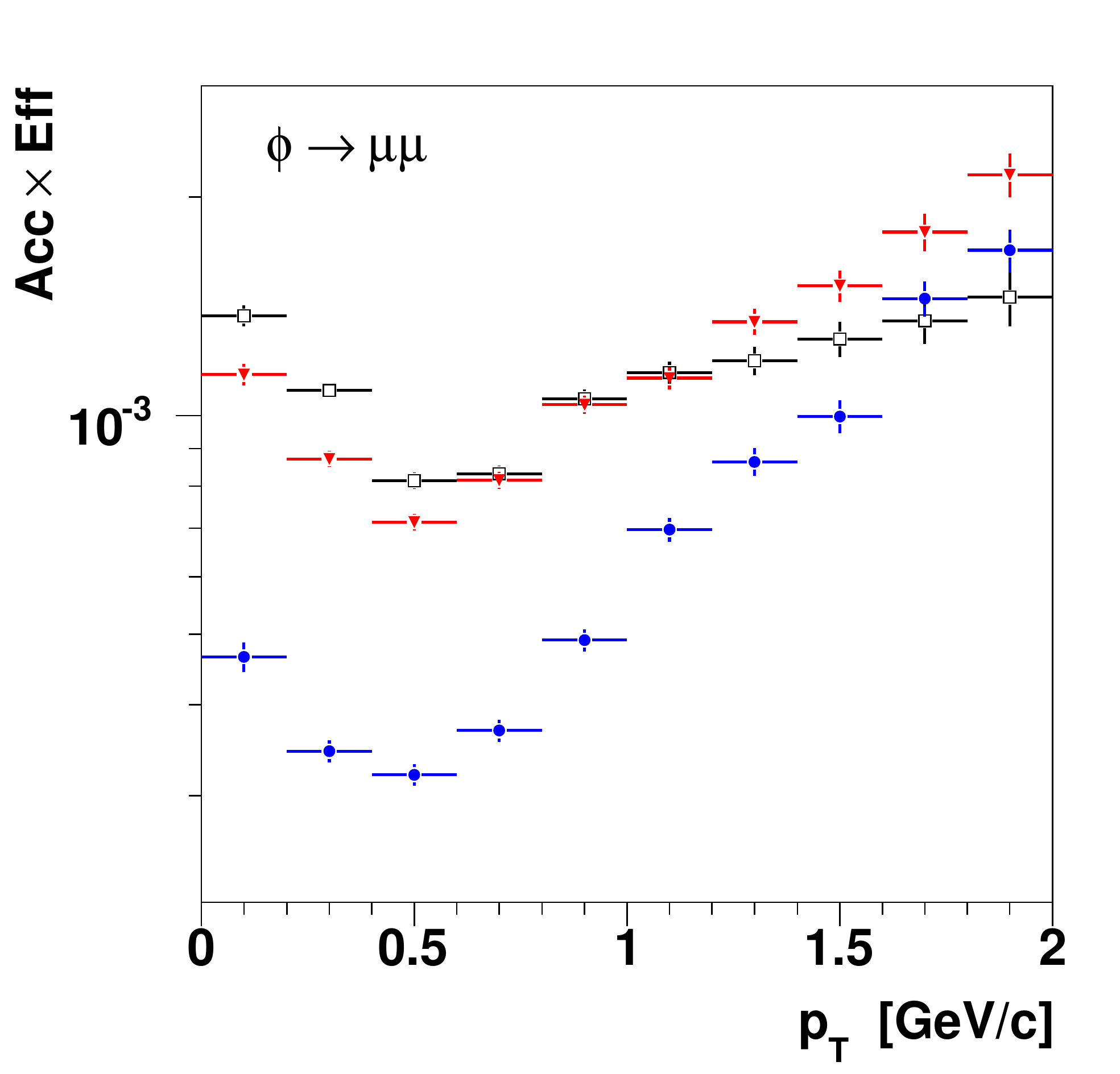}
    \end{center} 
\caption[\textwidth]{Acceptance $\times$ reconstruction efficiency vs $\pt$ for the
Uranium, Indium and Lead targets, for the three processes $\eta \to
\mu^+\mu^-\gamma$, $\omega \to \mu^+\mu^-$ and $\phi \to \mu^+\mu^-$.}
\label{fig:acceptance_vs_pt}
\end{figure*}
In \figurename~\ref{fig:acceptance_vs_pt} the dimuon acceptance $\times$ reconstruction
efficiency as a function of $\pt$ is
shown for the three processes $\eta\to\mu^+\mu^-\gamma$,
$\omega\to\mu^+\mu^-$ and $\phi\to\mu^+\mu^-$, on which the extraction of the
$\eta$, $\omega$ and $\phi$ yields is based. For each process, the
$\pt$ dependence is shown for the Uranium, Indium and Lead targets,
respectively sitting in the initial, central and final part of the
target system. In the 2004 p-A run the acceptance was lower than during the  In-In
run. This was due to a combination of several factors, the most
important being the replacement of 40 cm of graphite with 40 cm of
iron in the final part of the hadron absorber.
Other factors contributing to the loss of acceptance were the reduced
tracking efficiencies both for the muon spectrometer (due to 5~MWPC
broken planes) and the vertex spectrometer (one tracking plane
permanently switched off and the others having much reduced
efficiencies with respect to the In-In run, because of ageing effects),
as well as the selections imposed in the analysis for the
identification of the production target.

\section{Signal Extraction}

\noindent The signal extraction for the dimuon sources of interest for the present analysis
is based on a fit of the dimuon mass spectrum, 
after the subtraction of the combinatorial background. In the fit procedure, 
the data points are compared to 
the superposition of the expected MC sources in the mass region 
from the threshold up to 1.4~GeV/$c^2$, as explained
in~\cite{Arnaldi:2016pzu}.

Any possible $\rho/\omega$ interference effect is neglected in the fits performed in the present analysis, 
as justified by the results discussed in~\cite{Arnaldi:2016pzu}. 
The contribution of the Dalitz decay $\eta'\to
\mu^+\mu^-\gamma$ accounts for a very small fraction of the total
dimuon yield; for this reason, and because of its continuum shape
having no dominant structure apart from the broad peak at the $\rho$
mass (due to the contribution of the $\rho$ to the $\eta'$ form
factor), the fit to the reconstructed mass spectrum is not sensitive to
this contribution, and the ratio $\sigma_{\eta'} /
\sigma_\omega$ was fixed to 0.12~\cite{genesis,Becattini:2003wp}. 
The relative branching ratios $BR(\eta
\to\mu\mu)/BR(\eta\to\mu\mu\gamma)$ and $BR(\omega\to
\mu\mu\pi^0)/BR(\omega\to\mu\mu)$ have been fixed to the PDG value~\cite{Nakamura:2010zzi} and
to the value obtained from the analysis of the target-integrated mass spectrum~\cite{Arnaldi:2016pzu},
respectively. The normalisation of the $\rho$-meson 2-body decay has been left free in the $\pt$-integrated fits, 
where the available data samples were large enough to allow for a robust extraction of the corresponding signal; it was
otherwise fixed to $\sigma_\rho/\sigma_\omega = 1$, as justified by the $\pt$-integrated results.
All the other processes have their normalisations free.

The evaluation of the systematics on the signal extraction is performed by repeating
the fits on the dimuon mass spectra, each time varying the following
parameters of the input configuration: (i) the cross section ratio
$\sigma_{\eta'}/\sigma_{\omega}$ was varied by $\pm50\%$ with respect
to the nominal value~0.12; (ii) the combinatorial background
normalisation was varied by $\pm20\%$; (iii) the ratio $\sigma_\rho/\sigma_\omega$, when fixed,
was varied by $\pm10\%$ around the nominal value $\sigma_\rho/\sigma_\omega = 1$; (iv) the
relative branching ratios between the 2-body and Dalitz decays of the $\eta$ and $\omega$~mesons 
were varied by the uncertainties associated to the existing measurements~\cite{Nakamura:2010zzi,Arnaldi:2016pzu}; 
(v) the cut on the matching $\chi^2$ for the single muons was varied in the interval 2
to~3; (vi) two settings for the choice of the production vertices in the overlay~MC were used,
as described in the previous section. 

A possible bias in the estimation of the open charm contribution has also been considered, 
due to the fact that the Drell-Yan process, which does not give any appreciable
contribution below 1~GeV/$c^2$ while contributing above, is neglected in the fits.
In order to study the corresponding systematic effect, each fit has been repeated scaling the
open charm process down to 80\% and 60\% of the level optimised by
the fit when the contribution is left free.

\section{$\pt$ spectra of the $\rho/\omega$ and $\phi$ mesons}

\noindent In this section we present the analysis of the transverse
momentum spectra for the $\rho/\omega$ and $\phi$ mesons. 
To build the raw $\pt$ spectra we start by dividing the real data sample, target by target, into $\pt$ intervals of
200~MeV/$c$, sufficiently larger than the $\pt$ resolution of the apparatus in order for residual
smearing effects to be safely neglected. 
The same selection is then applied to the MC spectra for all
the signal sources, as well as to the contribution accounting for the
combinatorial background.  The available statistics is significant up to $\pt\sim 2$~GeV/$c$. For each target and $\pt$ interval, a fit is
then performed on the dimuon mass spectrum after the subtraction of the combinatorial background. In this way, the
contributions coming from the 2-body decays of the $\omega$ and $\phi$ mesons can be evaluated, and a raw $\pt$ spectrum is
extracted for each particle. This is then corrected for the acceptance
$\times$ efficiency as a function of $\pt$, estimated through the MC simulations. For
sufficiently narrow $\pt$ bins, this correction is independent of the
$\pt$ function used as input for the MC simulations.

Dividing the total sample into several $\pt$ intervals strongly reduces the statistics
available for each mass spectrum fit, making it impossible to properly disentangle 
the $\rho$ contribution under the $\omega$ peak. For
this reason, the constraint $\sigma_\rho/\sigma_\omega=1$ is imposed in these fits, as justified by the  
$\pt$-integrated analysis of the production cross section ratios as a function of the target size~A, discussed in the last section.
Whenever the $\sigma_\rho/\sigma_\omega=1$ constraint is imposed in the analysis, the results are referred to 
$\rho/\omega$ instead of $\rho$ or $\omega$ separately.

\begin{figure*}[p] 
   \begin{center}
    \vspace{0.5cm}
    \includegraphics[width=0.44\textwidth]{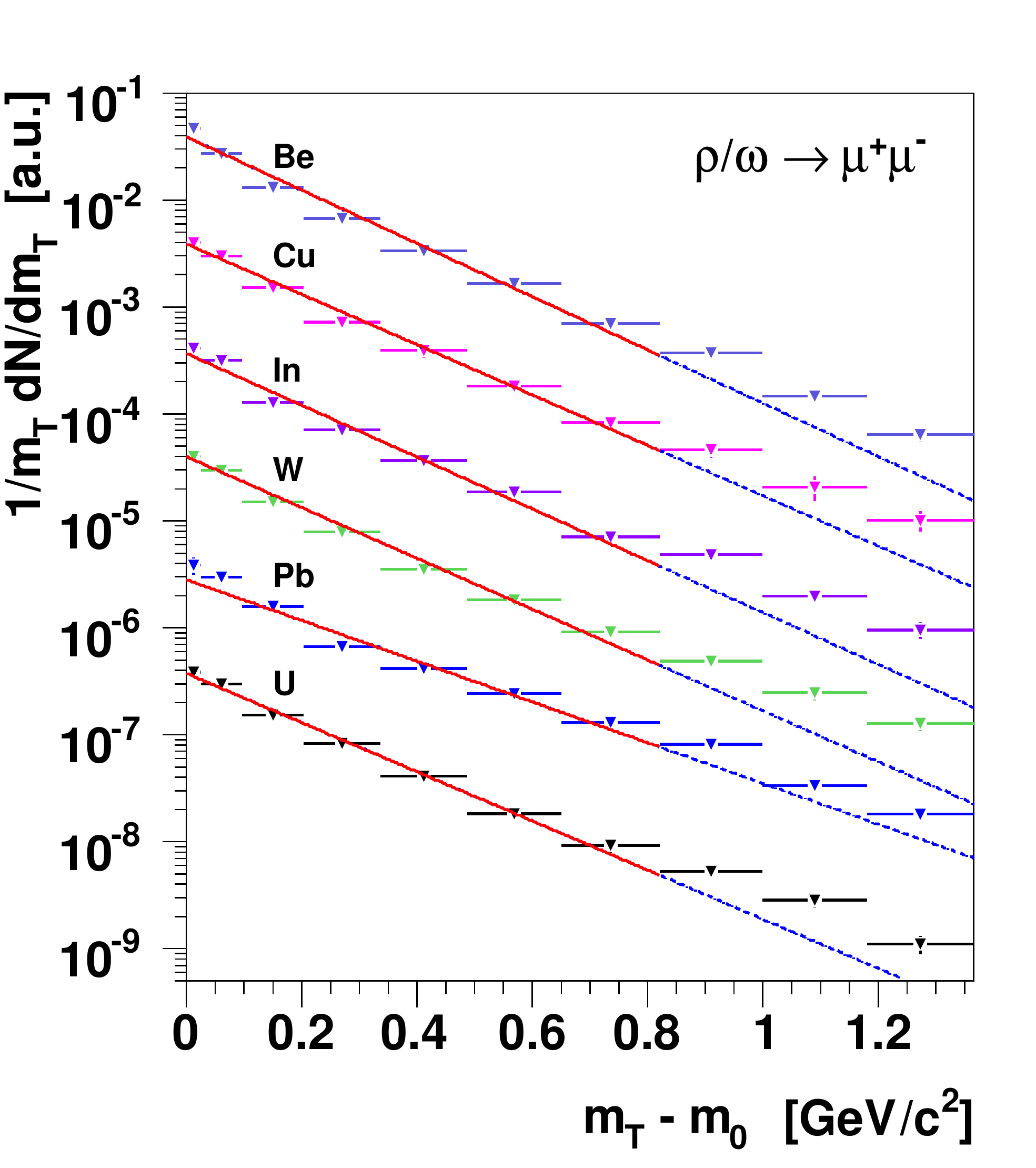} \hspace{.07\textwidth} 
    \includegraphics[width=0.44\textwidth]{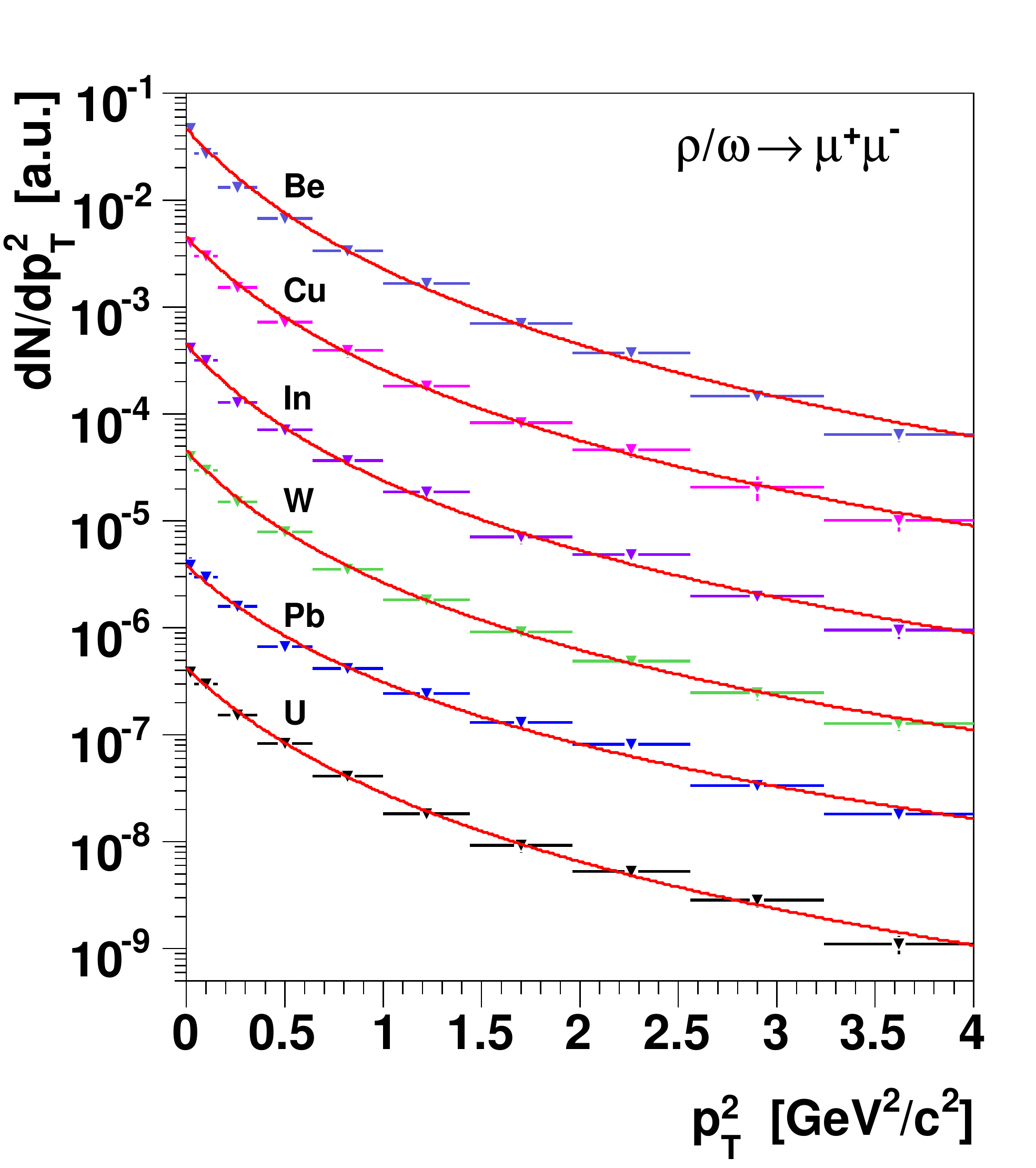} 
    \end{center} 
\caption[\textwidth]{Left: fits on the acceptance-corrected $\mt$
spectra of the $\rho/\omega$ mesons, with the exponential function~(\ref{eqn:thermal_pt_2}). Right: Fits on
the acceptance-corrected $\pt^2$ spectra of the $\rho/\omega$ mesons, with the power-law~(\ref{eqn:pt2_function_2}).}
\label{fig:mt_spectra_omega}
\end{figure*}

\begin{figure*}[p] 
   \begin{center}
    \vspace{1cm}
    \includegraphics[width=0.44\textwidth]{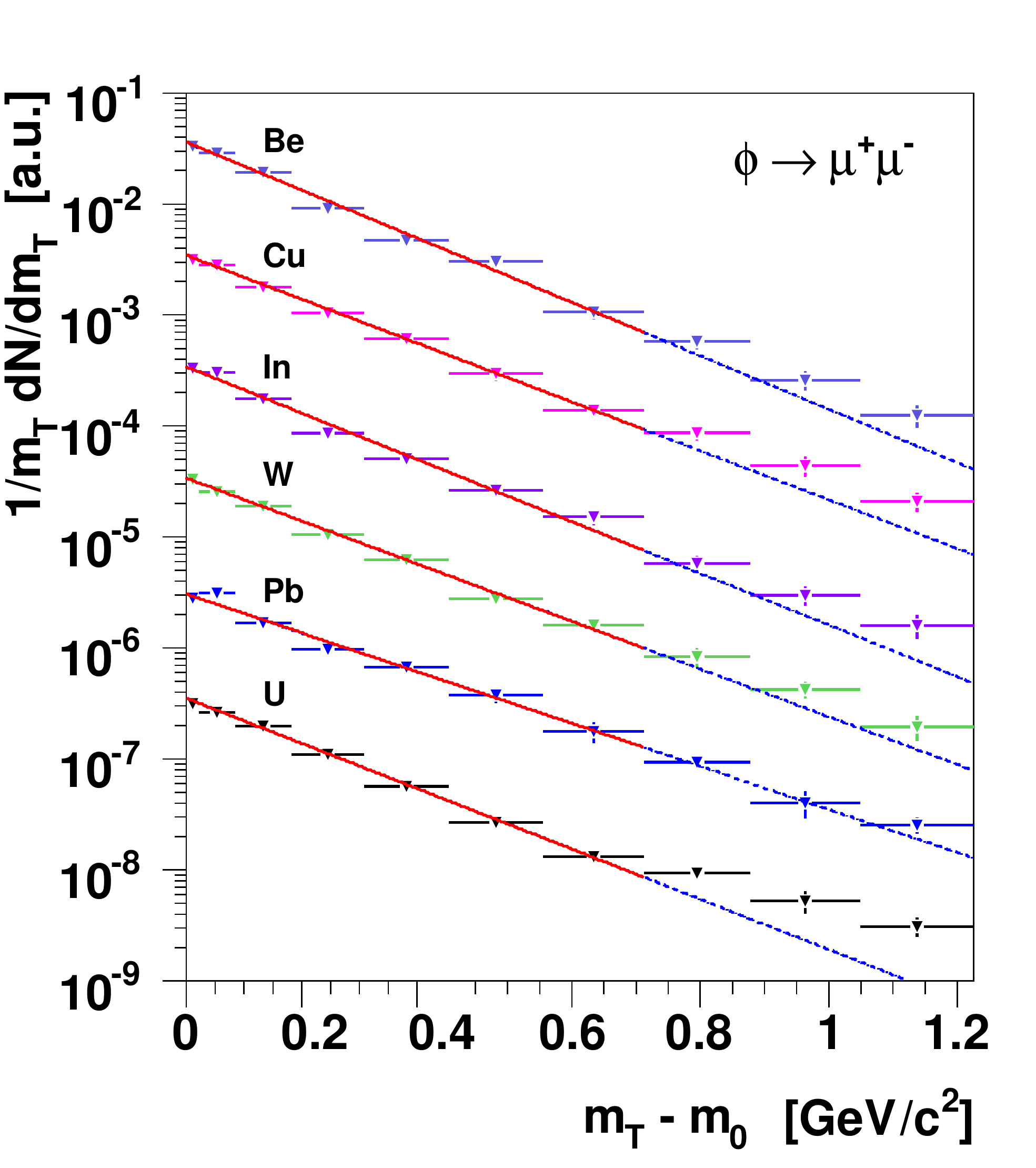} \hspace{.07\textwidth} 
    \includegraphics[width=0.44\textwidth]{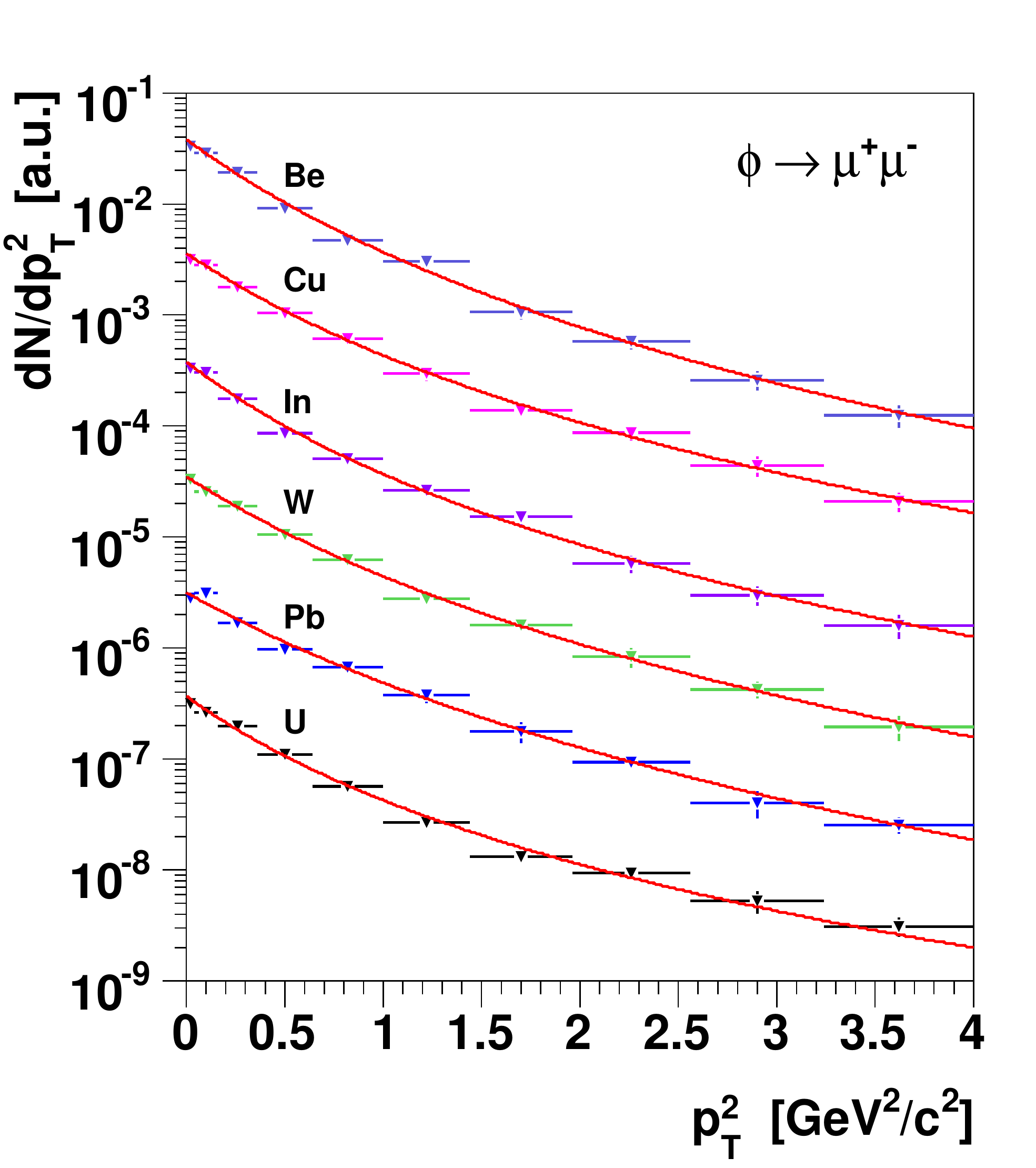}  
    \end{center} 
\caption[\textwidth]{Left: fits on the acceptance-corrected $\mt$
spectra of the $\phi$ meson, with the exponential function~(\ref{eqn:thermal_pt_2}). 
Right: Fits on the acceptance-corrected $\pt^2$
spectra of the $\phi$ meson, with the power-law~(\ref{eqn:pt2_function_2}).}
\label{fig:mt_spectra_phi}
\end{figure*}

The $\rho/\omega$ and $\phi$ $\pt$ spectra resulting after the correction
for the acceptance $\times$ efficiency are shown in
Figures~\ref{fig:mt_spectra_omega}--\ref{fig:mt_spectra_phi} as a
function of the transverse mass $\mt$ and $\pt^2$. The $\mt$ spectra
have been compared to the thermal-like function:
\begin{equation}
 \frac{1}{\pt} \frac{\d N}{\d \pt} = \frac{1}{\mt} \frac{\d N}{\d \mt} = 2
 \frac{\d N}{\d \pt^2} \propto \exp \left( -\frac{\mt}{T} \right)~.
\label{eqn:thermal_pt_2}
\end{equation} 
One can immediately appreciate how the thermal hypothesis clearly
fails in describing the spectra in the whole available kinematic range, showing a systematic
deviation from the pure exponential trend. The fact that such a deviation
was not observed in peripheral In-In collisions at 158~AGeV~\cite{Banicz:2009aa},
resulting in a hardening of the spectra reported in the present 
analysis, could be attributed to the larger collision energy in the p-A data.

\noindent The fit with the exponential function has thus been limited to
$(\mt - m_0) < 0.8$~GeV/$c^2$ and $(\mt - m_0) < 0.7$~GeV/$c^2$ 
for the $\rho/\omega$ and the $\phi$ mesons, respectively, 
where the thermal hypothesis
is found to describe the data reasonably well (solid, red line); the
extrapolation of the resulting fit function up to the upper end of the available
$\mt$ range is shown as a dashed, blue line, helping to appreciate
the contribution of the hard tail at high $\pt$. The $T$ values found
from the fits are compiled in \tablename~\ref{tab:values_pt_fit},
where both statistical and systematic uncertainties are reported, the 
latter reflecting squared sum of the systematic uncertainty on the signal extraction,
discussed in the previous Section. When evaluating the systematic uncertainty on the $T$~parameter, 
the upper limit of the fit range was also changed by the $\mt$ quantity equivalent to $200$ MeV/$c$ in $\pt$,
since the limit of the range where the thermal-like part of the distribution dominates
over the hard tail cannot be precisely established. \\

\noindent The superposition of a thermal-like distribution at low
$\pt$ and a hard tail at high $\pt$, favored by the data, can be modeled by means of the
following power-law function, used for example by the HERA-B Collaboration to
describe the $\pt^2$ spectrum of the $\phi$ meson~\cite{Abt:2006wt}:
\begin{equation}
 \frac{\d N}{\d \pt^2} \propto \left( 1 + \frac{\pt^2}{p_0^2} \right)^{-\beta}~,
 \label{eqn:pt2_function_2}
\end{equation}
a standard form where the $(1 + \ldots)$ term dominates in the limit of vanishing $\pt$, 
and the $\pt^{-\beta}$ dependence is characteristic of hard processes.  
The fits based on this power-law function successfully extend up to
the upper end of the available $\pt^2$ range, as it can be appreciated from the right panels of
\figurename~\ref{fig:mt_spectra_omega}--\ref{fig:mt_spectra_phi}.
The $\beta$ and $p_0^2$ parameters are listed in
\tablename~\ref{tab:values_pt_fit} with their statistical and
systematic uncertainties, the latter reflecting the squared sum of the
systematic uncertainties on the signal extraction,
discussed in the previous Section.
\begin{table*}[t]
  \vspace{0.2cm}
  \begin{center}

    \begin{tabular}{ c | l | c | c | c | c}

      \hline \hline

      \multicolumn{2}{r|}{\raisebox{-4pt}[0pt][10pt]{\textbf{Target~}}} & 
      \raisebox{-4pt}[0pt][10pt]{$\boldsymbol{T}$~\textbf{[MeV]}} & 
      \raisebox{-4pt}[0pt][10pt]{$\boldsymbol{p_0^2}$~~\textbf{[GeV}$\boldsymbol{^2/c^2}$\textbf{]}} &
      \raisebox{-4pt}[0pt][10pt]{$\boldsymbol{\beta}$} &
      \raisebox{-4pt}[0pt][10pt]{$\boldsymbol{\langle \pt \rangle~~[\mathrm{GeV}/c]}$}\\

      \hline \hline

      \rule{0.2cm}{0pt}

      \raisebox{-4pt}[0pt][10pt]{~} & 
      \raisebox{-4pt}[0pt][10pt]{Be} & 
      \raisebox{-4pt}[0pt][10pt]{$179 \pm 4 \pm 7$} & 
      \raisebox{-4pt}[0pt][10pt]{~~~$0.98 \pm 0.11 \pm 0.13$~~~} &
      \raisebox{-4pt}[0pt][10pt]{~~~$4.07 \pm 0.24 \pm 0.27$~~~} &
      \raisebox{-4pt}[0pt][10pt]{~~~~$0.57 \pm 0.04 \pm 0.05$~~~~}\\

      \raisebox{-4pt}[0pt][10pt]{~} & 
      \raisebox{-4pt}[0pt][10pt]{Cu} & 
      \raisebox{-4pt}[0pt][10pt]{$185 \pm 4 \pm 10$} & 
      \raisebox{-4pt}[0pt][10pt]{$0.79 \pm 0.10 \pm 0.12$} &
      \raisebox{-4pt}[0pt][10pt]{$3.45 \pm 0.22 \pm 0.24$} &
      \raisebox{-4pt}[0pt][10pt]{$0.60 \pm 0.05 \pm 0.06$}\\

      \raisebox{-4pt}[0pt][10pt]{~} & 
      \raisebox{-4pt}[0pt][10pt]{In} & 
      \raisebox{-4pt}[0pt][10pt]{$184 \pm 4 \pm 7$} & 
      \raisebox{-4pt}[0pt][10pt]{$0.69 \pm 0.09 \pm 0.08$} &
      \raisebox{-4pt}[0pt][10pt]{$3.27 \pm 0.20 \pm 0.15$} &
      \raisebox{-4pt}[0pt][10pt]{$0.62 \pm 0.05 \pm 0.03$}\\

      \raisebox{-4pt}[0pt][10pt]{$\rho/\omega$} & 
      \raisebox{-4pt}[0pt][10pt]{W} & 
      \raisebox{-4pt}[0pt][10pt]{$185 \pm 4 \pm 11$} & 
      \raisebox{-4pt}[0pt][10pt]{$0.66 \pm 0.07 \pm 0.02$} &
      \raisebox{-4pt}[0pt][10pt]{$3.08 \pm 0.16 \pm 0.12$} &
      \raisebox{-4pt}[0pt][10pt]{$0.62 \pm 0.05 \pm 0.03$}\\

      \raisebox{-4pt}[0pt][10pt]{~} & 
      \raisebox{-4pt}[0pt][10pt]{Pb} & 
      \raisebox{-4pt}[0pt][10pt]{$223 \pm 10 \pm 22$} & 
      \raisebox{-4pt}[0pt][10pt]{$0.74 \pm 0.14 \pm 0.08$} &
      \raisebox{-4pt}[0pt][10pt]{$2.94 \pm 0.24 \pm 0.17$} &
      \raisebox{-4pt}[0pt][10pt]{$0.69 \pm 0.10 \pm 0.06$}\\

      \raisebox{-4pt}[0pt][10pt]{~} & 
      \raisebox{-4pt}[0pt][10pt]{U} & 
      \raisebox{-4pt}[0pt][10pt]{$188 \pm 4 \pm 10$} & 
      \raisebox{-4pt}[0pt][10pt]{$0.82 \pm 0.10 \pm 0.10$} &
      \raisebox{-4pt}[0pt][10pt]{$3.40 \pm 0.21 \pm 0.23$} &
      \raisebox{-4pt}[0pt][10pt]{$0.62 \pm 0.05 \pm 0.06$}\\

      \raisebox{-4pt}[0pt][10pt]{~} & 
      \raisebox{-4pt}[0pt][10pt]{\textbf{ALL}} & 
      \raisebox{-4pt}[0pt][10pt]{$\boldsymbol{188 \pm 2 \pm 9}$} & 
      \raisebox{-4pt}[0pt][10pt]{$\boldsymbol{0.76 \pm 0.03 \pm 0.05}$} &
      \raisebox{-4pt}[0pt][10pt]{$\boldsymbol{3.31 \pm 0.07 \pm 0.13}$} &
      \raisebox{-4pt}[0pt][10pt]{$\boldsymbol{0.61 \pm 0.03 \pm 0.03}$}\\

      \hline

      \rule{0.2cm}{0pt}

      \raisebox{-4pt}[0pt][10pt]{~} & 
      \raisebox{-4pt}[0pt][10pt]{Be} & 
      \raisebox{-4pt}[0pt][10pt]{$183 \pm 6 \pm 6$} & 
      \raisebox{-4pt}[0pt][10pt]{$1.83 \pm 0.42 \pm 0.23$} &
      \raisebox{-4pt}[0pt][10pt]{$5.27 \pm 0.83 \pm 0.46$} &
      \raisebox{-4pt}[0pt][10pt]{~~~~$0.64 \pm 0.11 \pm 0.06$~~~~}\\

      \raisebox{-4pt}[0pt][10pt]{~} & 
      \raisebox{-4pt}[0pt][10pt]{Cu} & 
      \raisebox{-4pt}[0pt][10pt]{$198 \pm 6 \pm 6$} & 
      \raisebox{-4pt}[0pt][10pt]{$1.50 \pm 0.31 \pm 0.09$} &
      \raisebox{-4pt}[0pt][10pt]{$4.13 \pm 0.55 \pm 0.16$} &
      \raisebox{-4pt}[0pt][10pt]{$0.70 \pm 0.11 \pm 0.03$}\\

      \raisebox{-4pt}[0pt][10pt]{~} & 
      \raisebox{-4pt}[0pt][10pt]{In} & 
      \raisebox{-4pt}[0pt][10pt]{$188 \pm 6 \pm 13$} & 
      \raisebox{-4pt}[0pt][10pt]{$1.21 \pm 0.22 \pm 0.10$} &
      \raisebox{-4pt}[0pt][10pt]{$3.90 \pm 0.43 \pm 0.22$} &
      \raisebox{-4pt}[0pt][10pt]{$0.66 \pm 0.09 \pm 0.04$}\\

      \raisebox{-4pt}[0pt][10pt]{$\phi$} & 
      \raisebox{-4pt}[0pt][10pt]{W} & 
      \raisebox{-4pt}[0pt][10pt]{$202 \pm 6 \pm 5$} & 
      \raisebox{-4pt}[0pt][10pt]{$1.73 \pm 0.33 \pm 0.08$} &
      \raisebox{-4pt}[0pt][10pt]{$4.50 \pm 0.56 \pm 0.19$} &
      \raisebox{-4pt}[0pt][10pt]{$0.70 \pm 0.10 \pm 0.03$}\\

      \raisebox{-4pt}[0pt][10pt]{~} & 
      \raisebox{-4pt}[0pt][10pt]{Pb} & 
      \raisebox{-4pt}[0pt][10pt]{$222 \pm 9 \pm 16$} & 
      \raisebox{-4pt}[0pt][10pt]{$2.11 \pm 0.58 \pm 0.43$} &
      \raisebox{-4pt}[0pt][10pt]{$4.90 \pm 0.90 \pm 0.61$} &
      \raisebox{-4pt}[0pt][10pt]{$0.72 \pm 0.15 \pm 0.10$}\\

      \raisebox{-4pt}[0pt][10pt]{~} & 
      \raisebox{-4pt}[0pt][10pt]{U} & 
      \raisebox{-4pt}[0pt][10pt]{$192 \pm 5 \pm 1$} & 
      \raisebox{-4pt}[0pt][10pt]{$1.14 \pm 0.21 \pm 0.19$} &
      \raisebox{-4pt}[0pt][10pt]{$3.44 \pm 0.39 \pm 0.41$} &
      \raisebox{-4pt}[0pt][10pt]{$0.73 \pm 0.11 \pm 0.11$}\\

      \raisebox{-4pt}[0pt][10pt]{~} & 
      \raisebox{-4pt}[0pt][10pt]{\textbf{ALL}} & 
      \raisebox{-4pt}[0pt][10pt]{$\boldsymbol{193 \pm 2 \pm 5}$} & 
      \raisebox{-4pt}[0pt][10pt]{$\boldsymbol{1.27 \pm 0.10 \pm 0.07}$} &
      \raisebox{-4pt}[0pt][10pt]{$\boldsymbol{3.81 \pm 0.17 \pm 0.07}$} &
      \raisebox{-4pt}[0pt][10pt]{$\boldsymbol{0.70 \pm 0.04 \pm 0.02}$}\\

      \hline \hline

    \end{tabular}
  \end{center}
  \caption[\textwidth]{Results for the $\rho/\omega$ and $\phi$ transverse
momentum spectra as a function of A: $T$-values as obtained from
thermal functions; $p^2_0$, $\beta$ and $\langle \pt \rangle$ as
obtained from the power-law fit functions.}
  \label{tab:values_pt_fit}
\end{table*}

No significant trend for the $p_0^2$ or $\beta$ parameters,
as a function of the production target, can be inferred from the fit
results within the statistical and systematic uncertainties. To get
more insight into this aspect, in~\tablename~\ref{tab:values_pt_fit} we
also compile the mean value of $\pt$ for each target, as extracted
from the corresponding power-law fit function. The errors associated
to each value of $\langle \pt \rangle$ reflect the statistical and
systematic uncertainties on the $p_0^2$ or $\beta$ parameters.  No
definite trend as a function of the target can be identified here,
too. To compare the $\rho/\omega$ and $\phi$ $\pt$ spectra, then, the
target-integrated values are considered --- profiting from a better statistical accuracy. 
One finds $\langle \pt \rangle_{\rho/\omega}^\text{NA60,
p-A} = 0.61 \pm 0.03~\mathrm{(stat.)} \pm
0.03~\mathrm{(syst.)}~\mathrm{GeV}/c$, $\langle \pt
\rangle_\phi^\text{NA60, p-A} = 0.70 \pm 0.04~\mathrm{(stat.)} \pm
0.02~\mathrm{(syst.)}~\mathrm{GeV}/c$.
~\\
\noindent The results obtained for the $\pt$ spectra of the $\rho/\omega$ and $\phi$
mesons can be compared to the available experimental results. The
NA27 Collaboration measured the $\eta$, $\rho$, $\omega$ and $\phi$
production in p-p collisions at $\sqrt{s} = 27.5$~GeV~\cite{AguilarBenitez:1991yy}. 
No data point is shown by the NA27 Collaboration for the $\pt$ spectra of the $\omega$
and $\phi$ mesons, for which only the fit results are given, relative
to the function $\d N/\d \pt^2 \propto \exp(-\delta \pt^2)$, with the quoted
values for the $\delta$ parameter being: $\delta_\omega = 2.25 \pm
0.16$~(GeV/$c$)$^{-2}$, $\delta_\phi = 2.98 \pm
0.35$~(GeV/$c$)$^{-2}$. From these parameters one finds
$\langle \pt
 \rangle_\omega^\text{NA27} = (0.591 \pm 0.021)~\mathrm{GeV}/c$ and 
$\langle \pt \rangle_\phi^\text{NA27} = (0.513 \pm
 0.030)~\mathrm{GeV}/c$.
As one can see, the $\langle \pt \rangle$ for the $\omega$ meson extracted
from the fit function of NA27 agrees with the estimate obtained in the
present analysis, while for the $\langle \pt \rangle$ of
the $\phi$ meson a difference of more than 4 (statistical)
standard deviations is found. Furthermore, the present NA60 results clearly indicate
that $\langle \pt \rangle_\phi > \langle \pt \rangle_{\rho/\omega}$, while the
NA27 results imply that $\langle \pt \rangle_\phi < \langle \pt
\rangle_\omega$.

~\\
\noindent Concerning the $\pt$ measurement for the $\phi$-meson, the present results can
be also compared to the ones obtained by the HERA-B
Collaboration~\cite{Abt:2006wt}, which measured $\phi$-meson production (in
the $K^+K^-$ channel) in p-C, p-Ti and p-W collisions at $\snn =
41.6$~GeV. Keeping in mind the different energies between the NA60 and
the HERA-B data, the $\pt$ distributions are well described in both
cases by the power-law function~(\ref{eqn:pt2_function_2}).  As in
the present analysis, no definite trend for the $\langle \pt \rangle$ as a
function of the target is found by HERA-B. The $p_0^2$, $\beta$ parameters and the mean
$\pt$ measured by HERA-B are compatible with the NA60 measurements
shown in~\tablename~\ref{tab:values_pt_fit}; in particular, for the
Tungsten target common to both the NA60 and HERA-B data, the following values are found:
$ p_0^2(\text{HERA-B}) = 1.65 \pm 0.14~~(\mathrm{GeV}/c)^2$,
$\beta(\text{HERA-B}) = 4.20 \pm 0.14$,
$\langle \pt \rangle_\phi^\text{HERA-B, p-W} = (0.72 \pm 0.09)~\mathrm{GeV}/c$, to be compared to the 
NA60 results listed in~\tablename~\ref{tab:values_pt_fit}.

\section{Nuclear dependence of the $\eta$, $\omega$ and $\phi$ production cross sections}

\subsection{$\pt$-integrated analysis}

\label{nuclearDepPtIntegrated}

\begin{figure*}[p] 
   \begin{center}
    \includegraphics[width=0.44\textwidth]{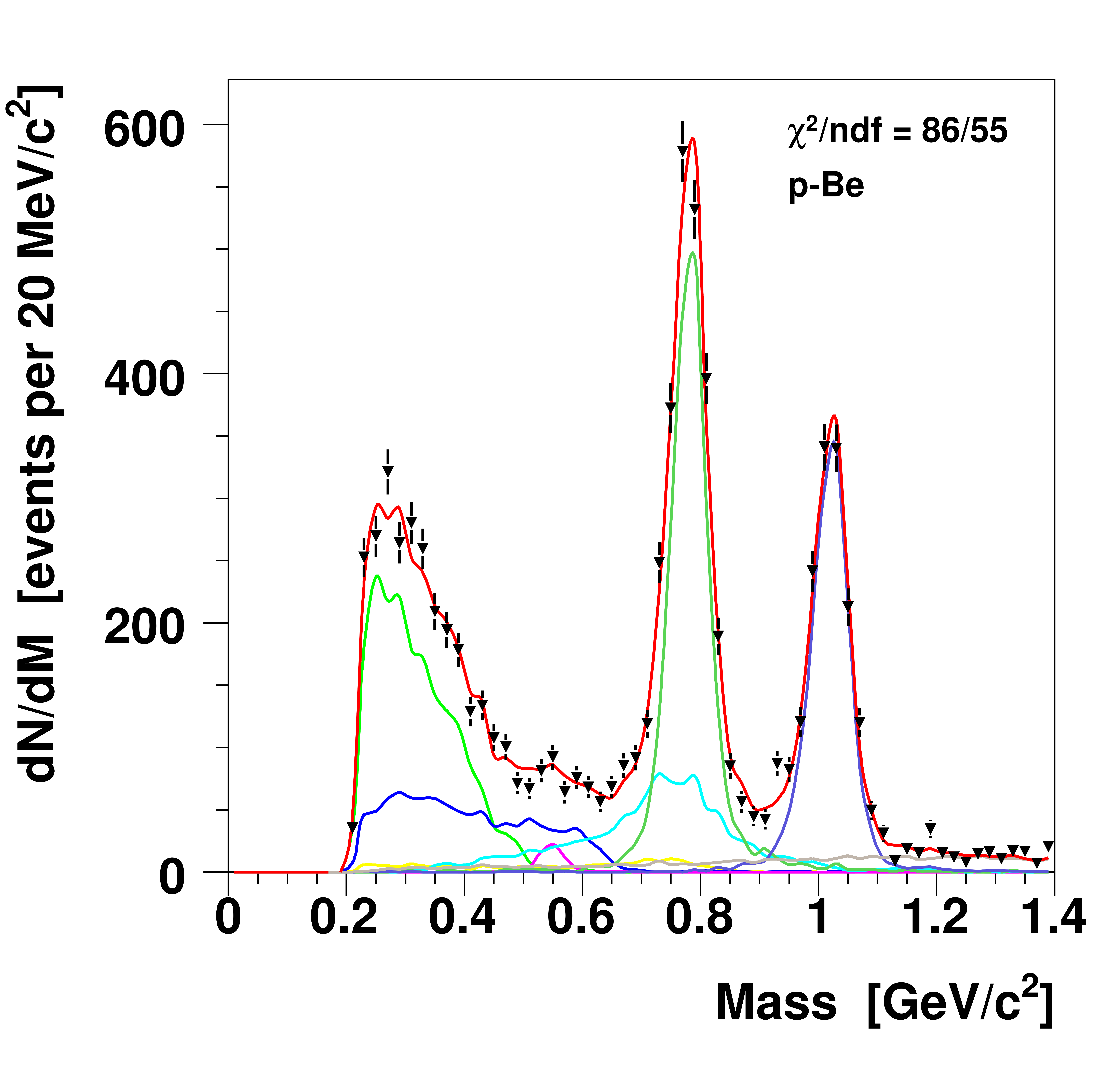} \hspace{.07\textwidth} 
    \includegraphics[width=0.44\textwidth]{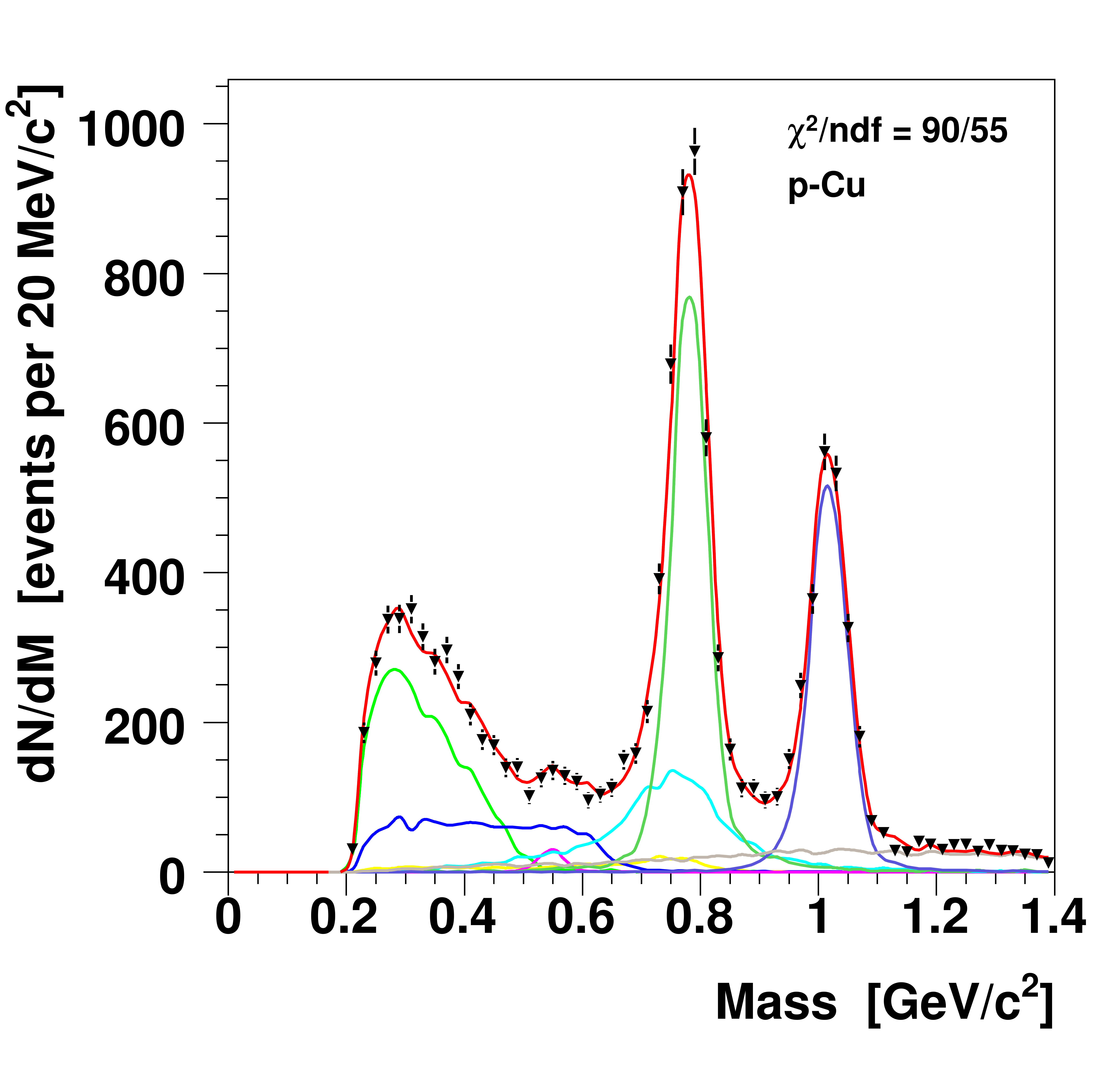} 
    \includegraphics[width=0.44\textwidth]{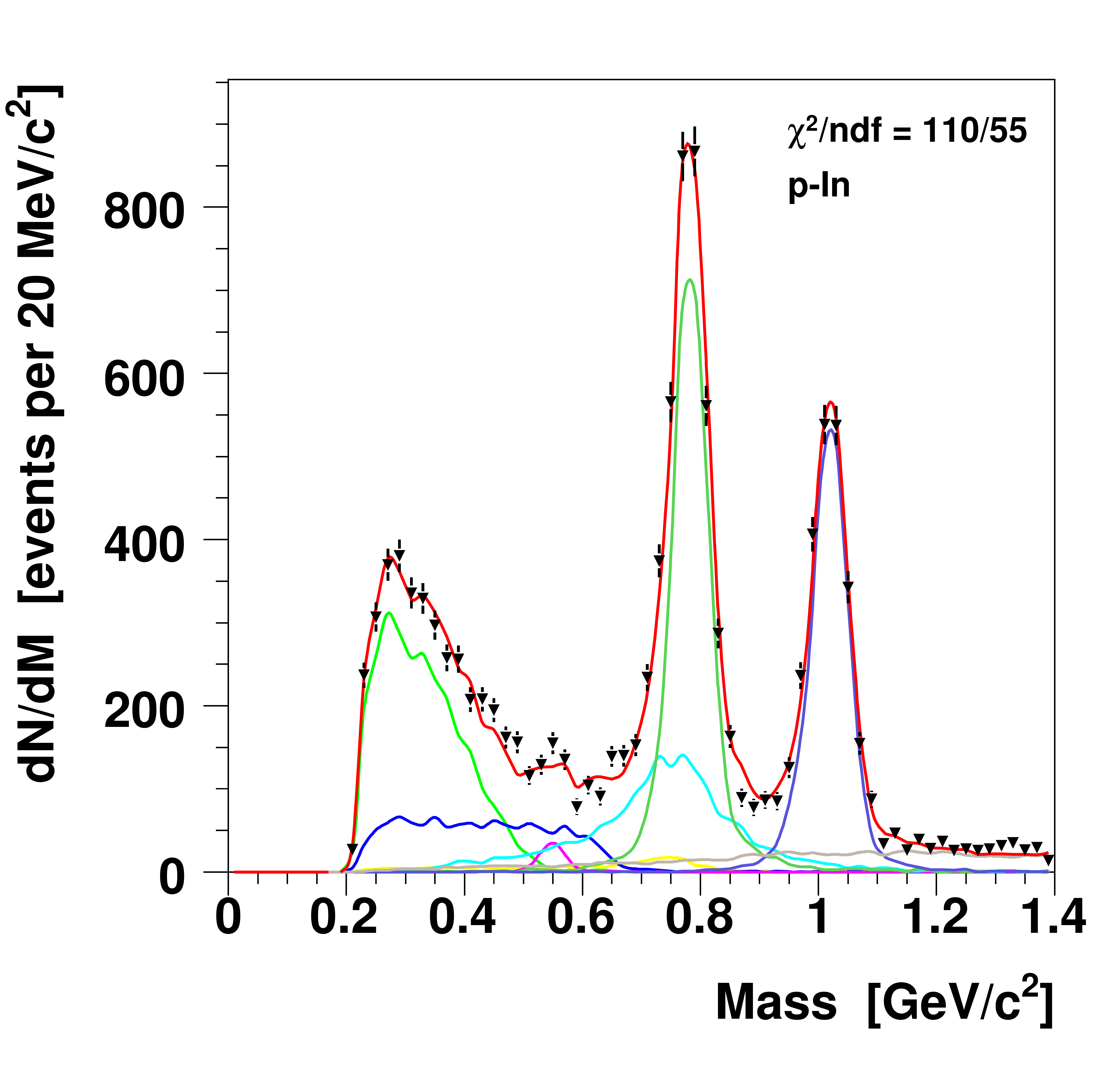} \hspace{.07\textwidth} 
    \includegraphics[width=0.44\textwidth]{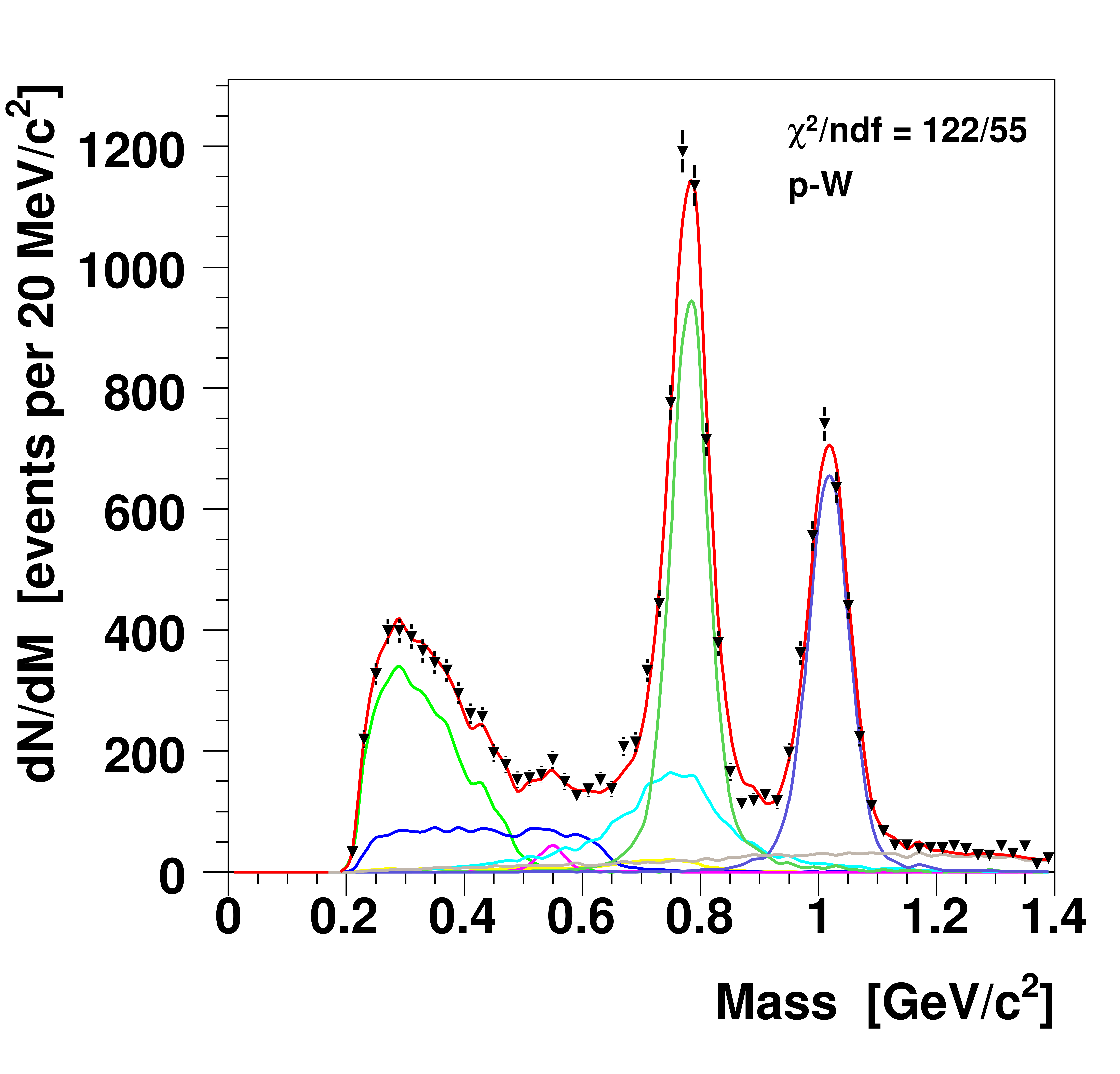} 
    \includegraphics[width=0.44\textwidth]{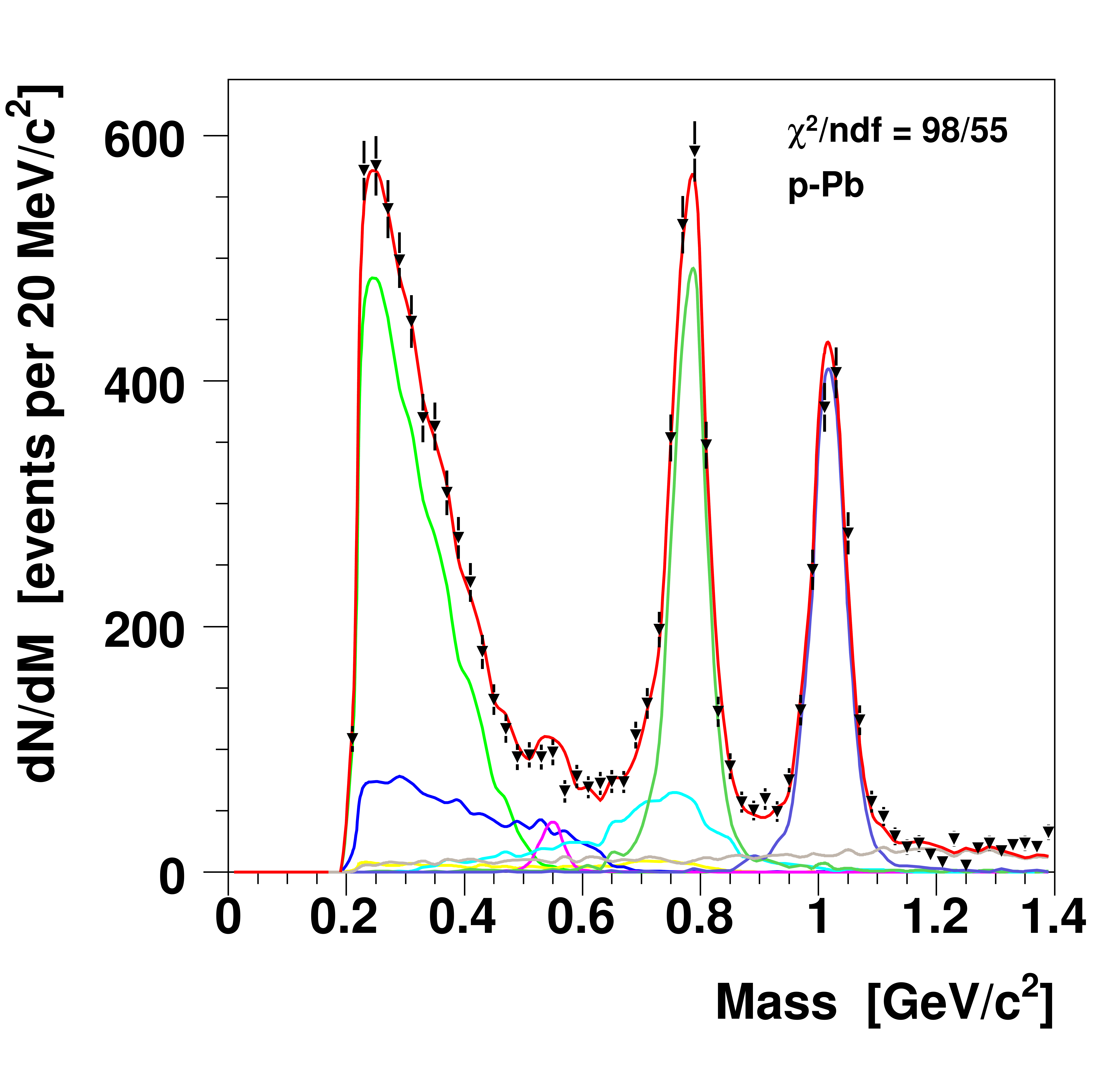} \hspace{.07\textwidth} 
    \includegraphics[width=0.44\textwidth]{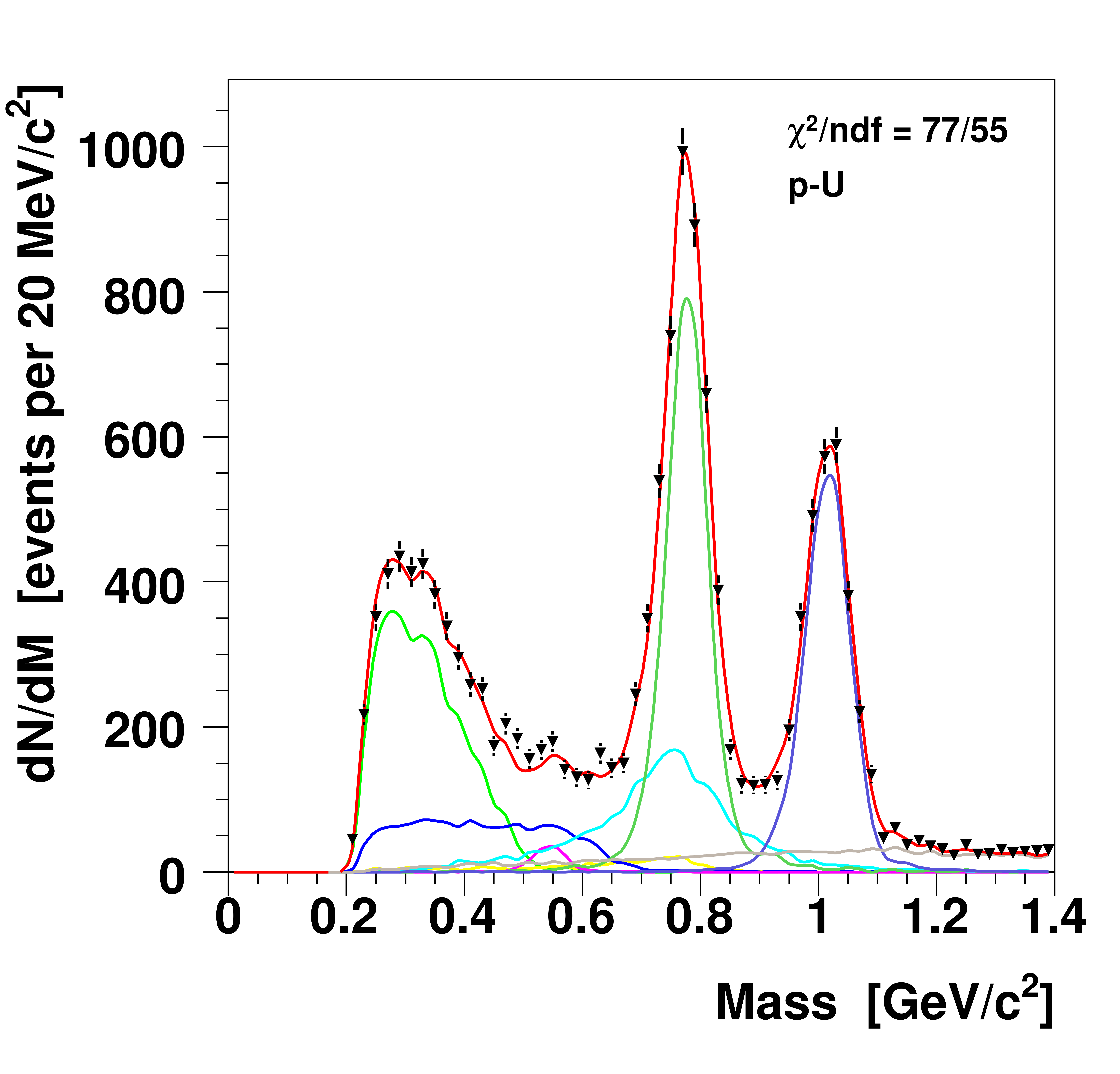} 
    \end{center} 
\caption[\textwidth]{Fits on the $\pt$-integrated dimuon mass spectra
for the different production targets, with the superposition of the expected sources: $\eta\to\mu^+\mu^-\gamma$ (light green, dominating for masses below 0.5~GeV/$c^2$),
$\eta\to\mu^+\mu^-$ (magenta), $\rho\to\mu^+\mu^-$ (cyan), $\omega\to\mu^+\mu^-\pi^0$ (blue), $\omega\to\mu^+\mu^-$ (dark green),
$\eta'\to\mu^+\mu^-\gamma$ (yellow), $\phi\to\mu^+\mu^-$ (dark violet), and the open charm process (grey).}
\label{fig:rawMassSpectra_ptIntegrated}
\end{figure*}

\noindent The $\pt$-integrated yields of the $\eta$, $\omega$ and $\phi$ mesons are extracted from the
fits shown in \figurename~\ref{fig:rawMassSpectra_ptIntegrated}.
As one can see, the comparison between the sum of the
MC sources and the dimuon mass spectrum is satisfactory over the whole
mass range for each target. The yields
are then corrected for the acceptance $\times$ efficiency
evaluated in the full phase space, and normalised to the
number of nuclear interaction lengths of each target. The small beam attenuation seen by the downstream targets
is also taken into account and corrected for. In the case of the $\eta$-meson, the difference between the kinematics of the parent
meson and the dimuon has been properly estimated by means of a MC simulation of the Dalitz process, and the results properly corrected for.

~\\
\noindent In the following, the nuclear dependence of the production cross-sections, 
normalised to the lightest target is presented. The absolute cross-sections could not be determined 
because of a malfunctioning of the argonium counter devoted the beam luminosity measurement.
The relative production cross sections for the $\omega$,
$\phi$ and $\eta$ mesons,
normalised to the Beryllium target, are shown in the left panels of
\figurename~\ref{fig:alphaVsPt}. The cross section for the $\rho$-meson was found to be
compatible with the one of the $\omega$-meson within the errors, independent of the target, and is not shown here (see Section~\ref{particleRatios}). 
The nuclear dependence of the production cross
sections has been parameterised with the power law $\sigma_\mathrm{pA} \propto
\mathrm{A}^\alpha$~\cite{Sibirtsev:2008ib}: the fit functions are shown as solid, red lines in
\figurename~\ref{fig:alphaVsPt}, giving $\alpha_\omega = 0.841 \pm 0.014~\mathrm{(stat.)} \pm
0.030~\mathrm{(syst.)}$, $\alpha_\phi = 0.906\pm
0.011~\mathrm{(stat.)} \pm 0.025~\mathrm{(syst.)}$ and 
$\alpha_\eta = 0.935 \pm 0.048~\mathrm{(stat.)} \pm 0.060~\mathrm{(syst.)}$.\\
\begin{table*}[t]
  \vspace{0.2cm}
  \begin{center}

    \begin{tabular}{ c | c | c | c }

      \hline \hline

      \raisebox{-4pt}[0pt][10pt]{$\boldsymbol{\mathrm{\Delta} \pt}$~\textbf{[GeV/$\boldsymbol{c}$]}} & 
      \raisebox{-4pt}[0pt][10pt]{$\boldsymbol{\alpha_{\eta}}$} &
      \raisebox{-4pt}[0pt][10pt]{$\boldsymbol{\alpha_{\omega}}$} &
      \raisebox{-4pt}[0pt][10pt]{$\boldsymbol{\alpha_{\phi}}$} \\

      \hline \hline

      \raisebox{-4pt}[0pt][10pt]{$0.0 - 0.2$} & 
      \raisebox{-4pt}[0pt][10pt]{---} &
      \raisebox{-4pt}[0pt][10pt]{~~~$0.77 \pm 0.03 \pm 0.01$~~~} &
      \raisebox{-4pt}[0pt][10pt]{~~~$0.87 \pm 0.04 \pm 0.02$~~~}\\

      \raisebox{-4pt}[0pt][10pt]{$0.2 - 0.4$} & 
      \raisebox{-4pt}[0pt][10pt]{---} &
      \raisebox{-4pt}[0pt][10pt]{~~~$0.77 \pm 0.02 \pm 0.02$~~~} &
      \raisebox{-4pt}[0pt][10pt]{~~~$0.85 \pm 0.02 \pm 0.01$~~~}\\

      \raisebox{-4pt}[0pt][10pt]{$0.4 - 0.6$} & 
      \raisebox{-4pt}[0pt][10pt]{---} &
      \raisebox{-4pt}[0pt][10pt]{~~~$0.82 \pm 0.02 \pm 0.02$~~~} &
      \raisebox{-4pt}[0pt][10pt]{~~~$0.86 \pm 0.03 \pm 0.02$~~~}\\

      \raisebox{-4pt}[0pt][10pt]{$0.6 - 0.8$} & 
      \raisebox{-4pt}[0pt][10pt]{~~~$0.85 \pm 0.04 \pm 0.04$~~~} &
      \raisebox{-4pt}[0pt][10pt]{~~~$0.80 \pm 0.02 \pm 0.01$~~~} &
      \raisebox{-4pt}[0pt][10pt]{~~~$0.90 \pm 0.03 \pm 0.02$~~~}\\

      \raisebox{-4pt}[0pt][10pt]{$0.8 - 1.0$} & 
      \raisebox{-4pt}[0pt][10pt]{~~~$0.91 \pm 0.04 \pm 0.02$~~~} &
      \raisebox{-4pt}[0pt][10pt]{~~~$0.80 \pm 0.02 \pm 0.02$~~~} &
      \raisebox{-4pt}[0pt][10pt]{~~~$0.93 \pm 0.03 \pm 0.03$~~~}\\

      \raisebox{-4pt}[0pt][10pt]{$1.0 - 1.2$} & 
      \raisebox{-4pt}[0pt][10pt]{~~~$0.92 \pm 0.04 \pm 0.02$~~~} &
      \raisebox{-4pt}[0pt][10pt]{~~~$0.82 \pm 0.02 \pm 0.03$~~~} &
      \raisebox{-4pt}[0pt][10pt]{~~~$0.92 \pm 0.03 \pm 0.04$~~~}\\

      \raisebox{-4pt}[0pt][10pt]{$1.2 - 1.4$} & 
      \raisebox{-4pt}[0pt][10pt]{~~~$0.98 \pm 0.03 \pm 0.02$~~~} &
      \raisebox{-4pt}[0pt][10pt]{~~~$0.89 \pm 0.03 \pm 0.01$~~~} &
      \raisebox{-4pt}[0pt][10pt]{~~~$0.98 \pm 0.04 \pm 0.02$~~~}\\

      \raisebox{-4pt}[0pt][10pt]{$1.4 - 1.6$} & 
      \raisebox{-4pt}[0pt][10pt]{~~~$1.01 \pm 0.04 \pm 0.02$~~~} &
      \raisebox{-4pt}[0pt][10pt]{~~~$0.91 \pm 0.03 \pm 0.04$~~~} &
      \raisebox{-4pt}[0pt][10pt]{~~~$1.02 \pm 0.06 \pm 0.04$~~~}\\

      \raisebox{-4pt}[0pt][10pt]{$1.6 - 1.8$} & 
      \raisebox{-4pt}[0pt][10pt]{~~~$0.96 \pm 0.05 \pm 0.04$~~~} &
      \raisebox{-4pt}[0pt][10pt]{~~~$0.97 \pm 0.03 \pm 0.01$~~~} &
      \raisebox{-4pt}[0pt][10pt]{~~~$1.04 \pm 0.07 \pm 0.04$~~~}\\

      \raisebox{-4pt}[0pt][10pt]{$1.8 - 2.0$} & 
      \raisebox{-4pt}[0pt][10pt]{~~~$0.98 \pm 0.08 \pm 0.04$~~~} &
      \raisebox{-4pt}[0pt][10pt]{~~~$0.98 \pm 0.05 \pm 0.02$~~~} &
      \raisebox{-4pt}[0pt][10pt]{~~~$1.09 \pm 0.08 \pm 0.02$~~~}\\

      \hline \hline

    \end{tabular}
  \end{center}
  \caption[\textwidth]{Results for the $\alpha$ parameter as a function
of $\pt$ for the $\eta$, $\omega$ and $\phi$ mesons.}
  \label{tab:values_alpha}
\end{table*}

\noindent Fixing the reference for the relative cross sections to the Beryllium nucleus 
could induce a bias in the analysis of the nuclear dependence of particle production. 
Accounting for 9 nucleons only, indeed, the Be nucleus
could behave more as an incoherent superposition of single nucleons
rather than as a nuclear system having collective properties --- as in
the case of the heavier nuclei. This could lead to a breaking
of the power-law dependence $\sigma_\mathrm{pA} \propto \mathrm{A}^\alpha$
for the Be target, biasing the extraction of the
$\alpha$ parameters. For this reason, the option to
normalise the production cross sections to the Cu target, excluding the Be 
point from the fit, was also considered as a cross-check: the corresponding results on the $\alpha$
parameters, affected by larger statistical uncertainties because of the much reduced lever arm, and not shown here,
were found to be compatible within the uncertainties to the ones presented above.

\subsection{$\pt$-differential analysis}

\begin{figure*}[p] 
\vspace{-0.5cm}
   \begin{center}
   \includegraphics[width=0.44\textwidth]{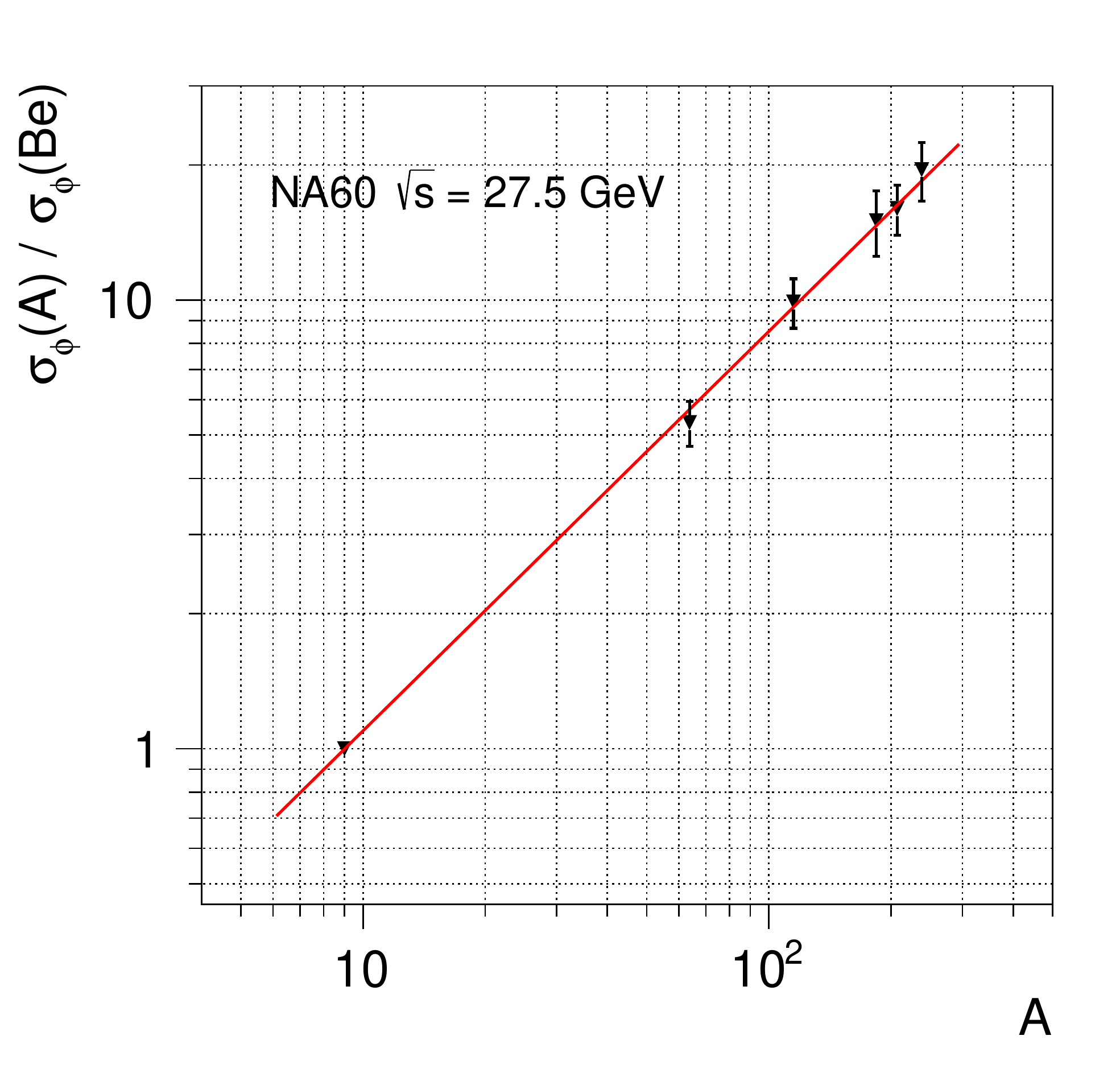}\hspace{0.04\textwidth}
    \includegraphics[width=0.44\textwidth]{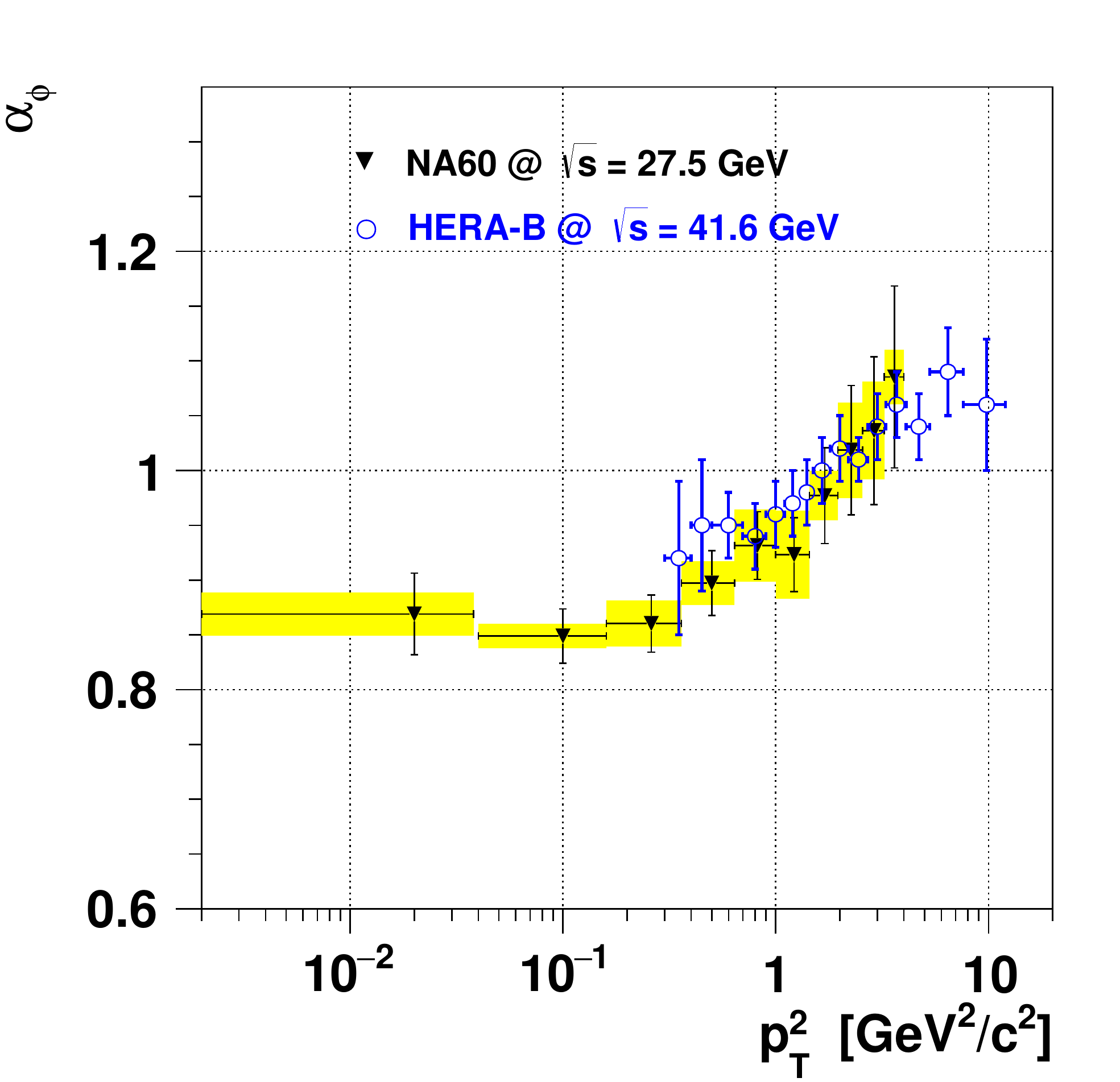}\\
    \includegraphics[width=0.44\textwidth]{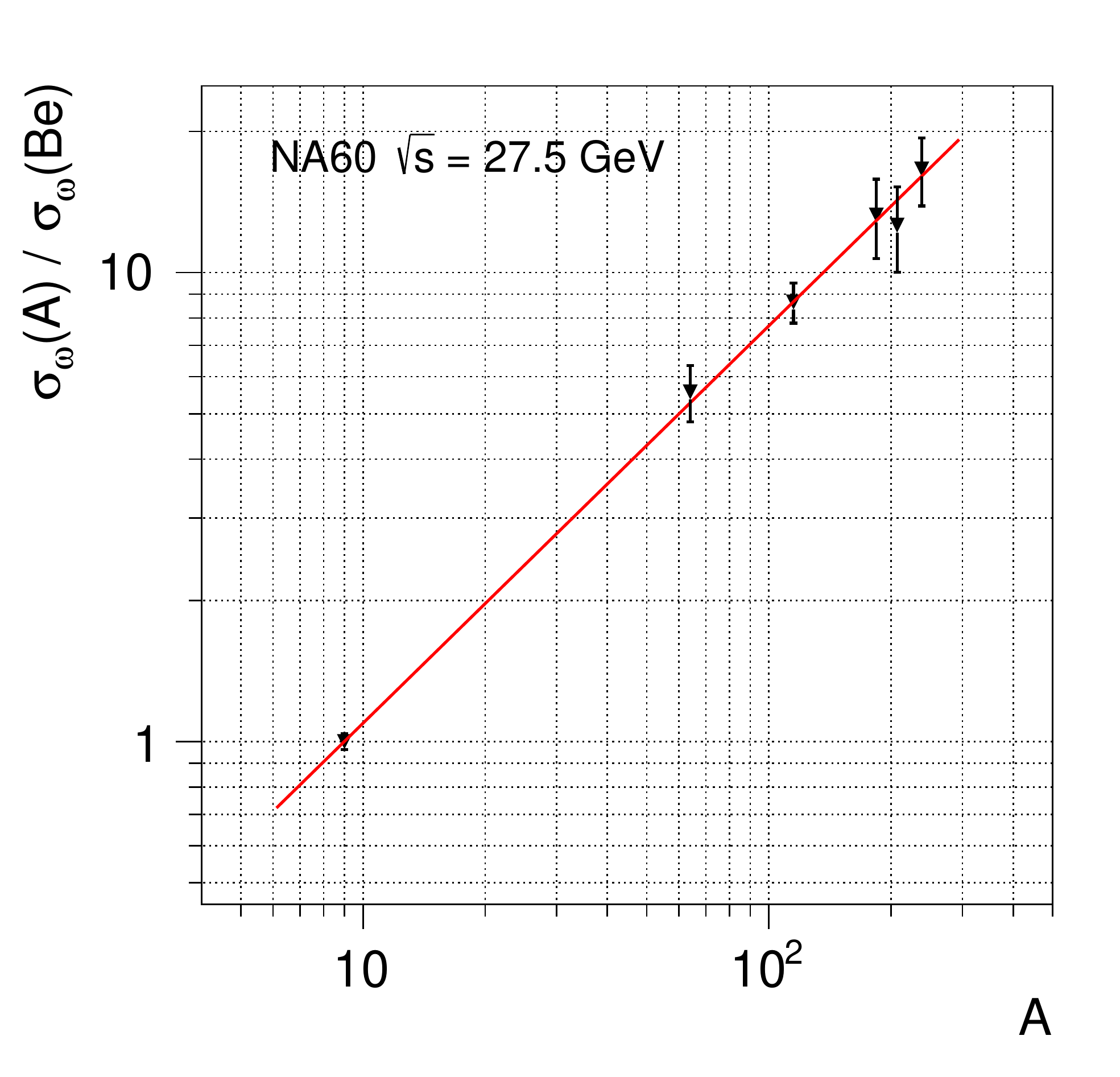}\hspace{0.04\textwidth}
    \includegraphics[width=0.44\textwidth]{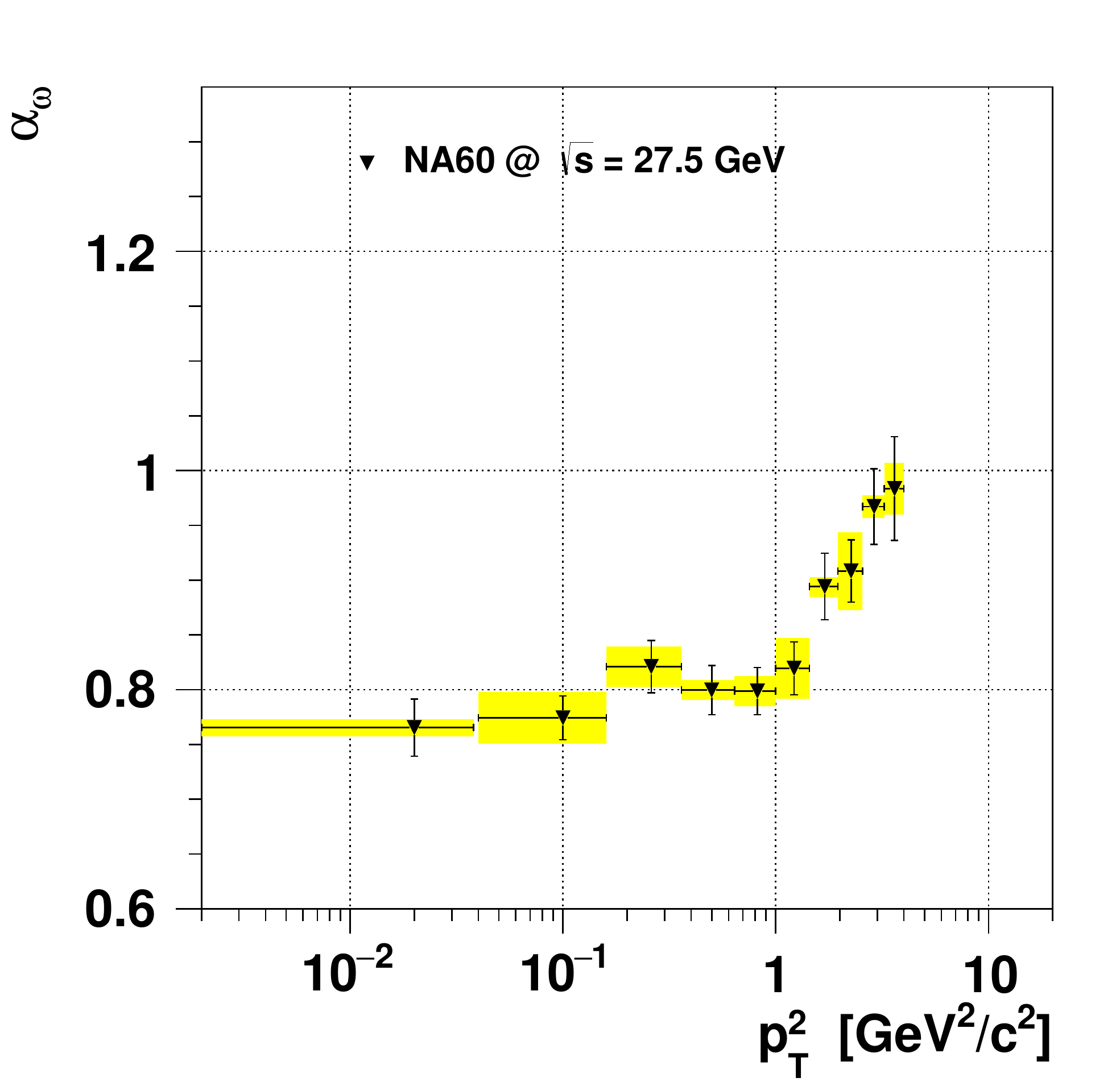} \\
    \includegraphics[width=0.44\textwidth]{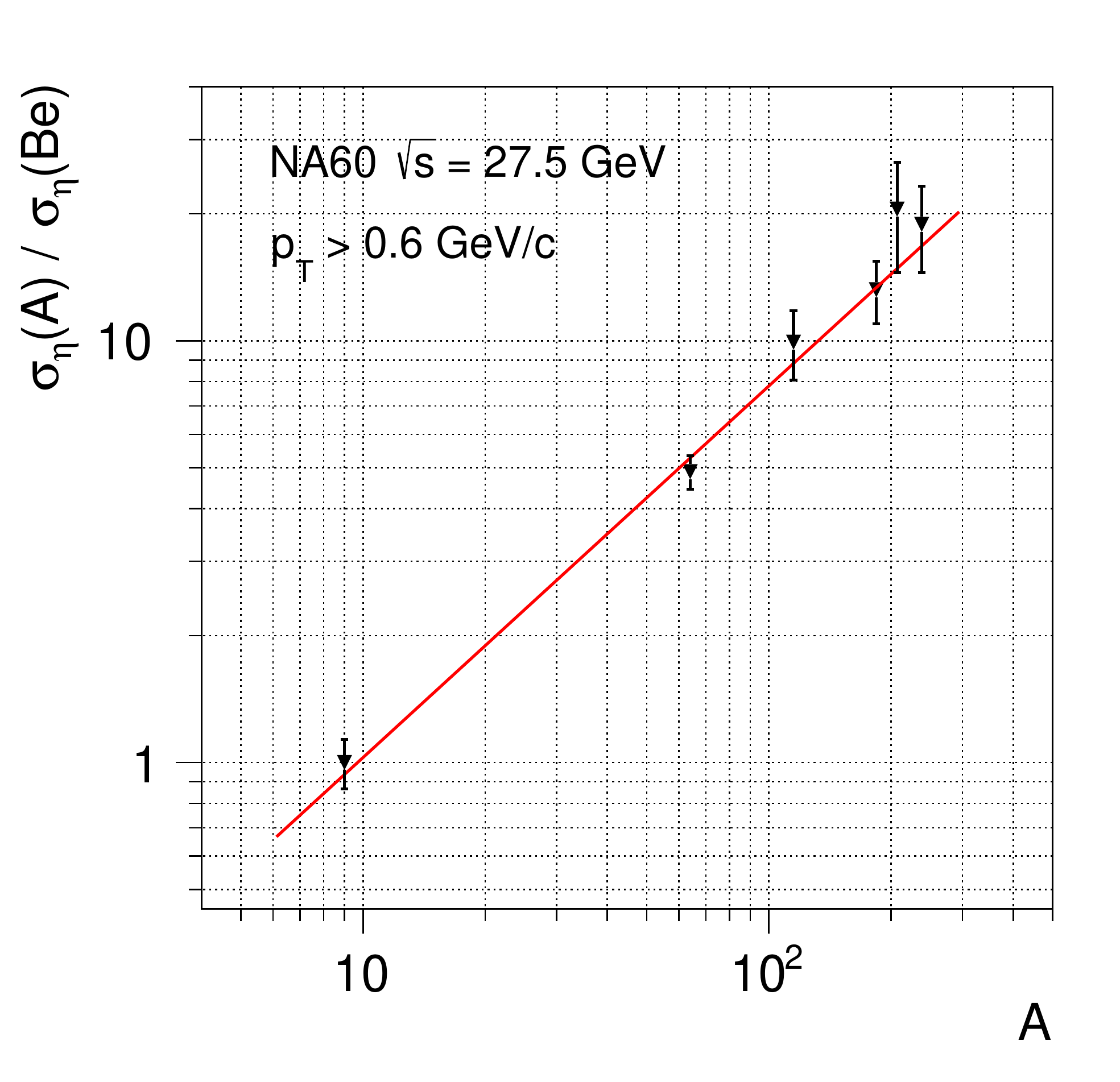}\hspace{0.04\textwidth}
    \includegraphics[width=0.44\textwidth]{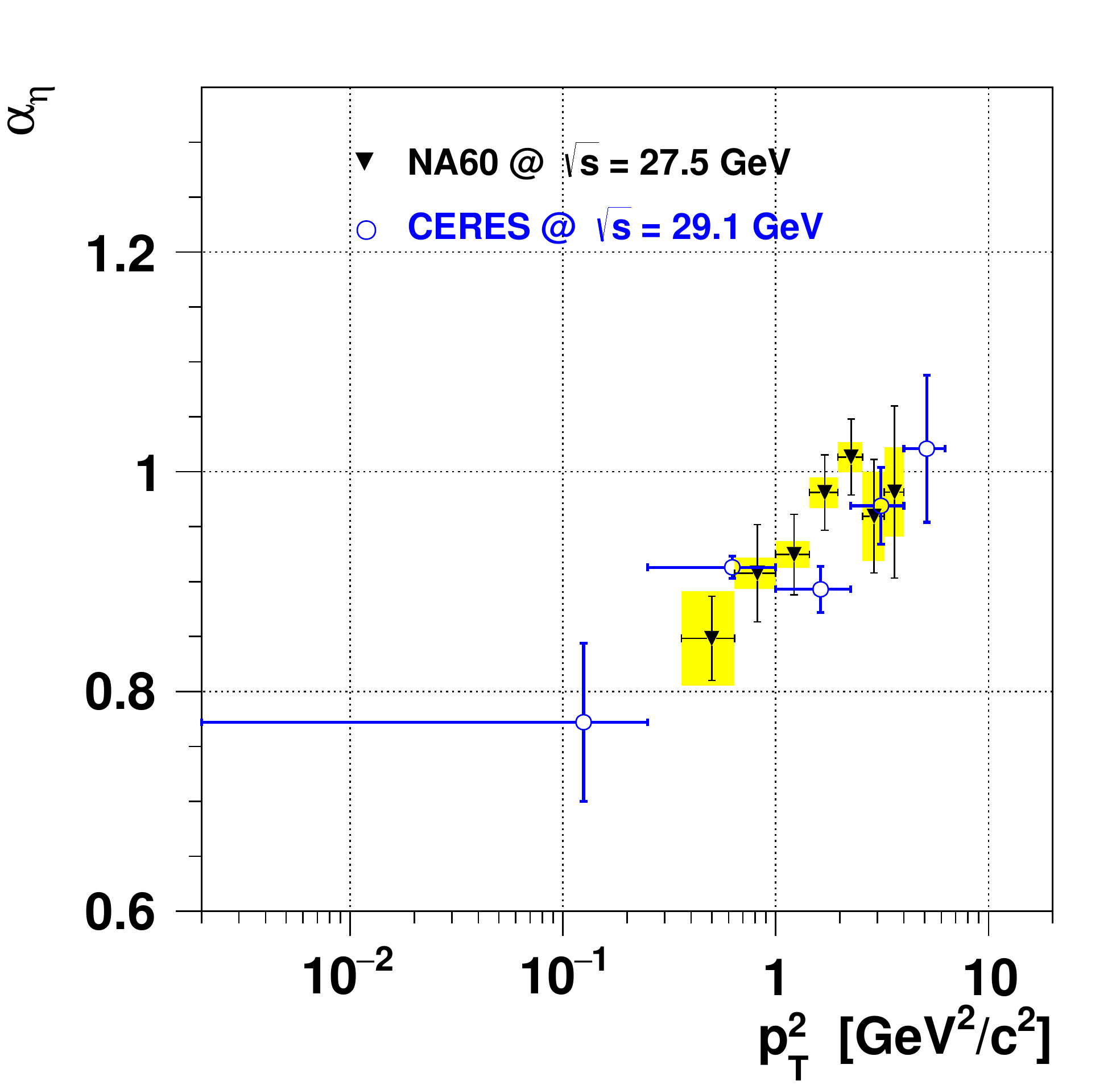}
    \end{center} 
\vspace{-0.3cm}
\caption[\textwidth]{Left column:
fit on the relative production cross sections for the $\omega$, $\phi$
and $\eta$ mesons as a function of A, normalized to the Be or the Cu
target. Right column: $\pt$ dependence of the $\alpha$ parameter for the $\omega$, $\phi$ and $\eta$ mesons.}
\label{fig:alphaVsPt}
\vspace{0.5cm}
\end{figure*}

\noindent The $\pt$-dependence of the $\alpha$ parameters for the $\eta$, $\omega$ and $\phi$
mesons has also been studied, considering $\pt$ intervals of 200~MeV/$c$ and starting from $\pt = 0$ for the
$\omega$ and $\phi$ mesons and from $\pt = 0.6$~GeV/$c$ for the $\eta$
meson. For each slice of $\pt$, the $\alpha_\eta$,
$\alpha_\omega$ and $\alpha_\phi$ parameters have been extracted, together
with their statistical and systematic uncertainties, by applying the same procedure
discussed in the previous section for the $\pt$-integrated measurement. 
The $\alpha$ parameters are shown in \tablename~\ref{tab:values_alpha} and their 
$\pt$ dependence is summarised on the right column of
\figurename~\ref{fig:alphaVsPt}: in each plot, the error bars and
the shadowed boxes account for the statistical and systematic
uncertainties, respectively. It was also verified (not shown here) that the same trend of the $\alpha$ parameters as a function 
of $\pt$, although with significantly larger uncertainties, could be found 
when excluding the Be target from the analysis, setting the relative normalisation to the Cu target.

~\\
\noindent The $\alpha$ parameters increase as a function of
$\pt$ for the three considered particles, a behaviour which can be related to the 
so-called ``Cronin effect'' originally
observed by Cronin~\emph{et al.}~for charged
kaons~\cite{Kluberg:1977bm}. The trend of the $\alpha$ parameters
can also be interpreted as a consequence of the hardening of the particle production mechanism with $\pt$,
with values of $\alpha$ closer to the ``black disk'' limit $\alpha = 2/3$ for $\pt \approx 0$, as expected for
soft production mechanisms scaling with the surface of the target nucleus,
and values closer to the $\alpha = 1$ limit for $\pt \gtrsim 1.5$~GeV/$c$, as expected for hard production mechanisms scaling with
the number of available nucleons.
An additional hardening of the production mechanism could be attributed to the $s\bar{s}$ 
component in the quark wave function of the measured hadrons: in absence of explicit theoretical calculations for the $\alpha$ parameters,
this simple assumption could still qualitatively explain the larger values measured for $\alpha_\phi$ and $\alpha_\eta$, 
with respect to $\alpha_\omega$.

~\\
\noindent The nuclear dependence of the production cross sections for
the light neutral mesons has been studied by
previous experiments for the $\phi$ meson, although at energy regimes often different from
the one considered in the present analysis.  The $\alpha$ parameter
for the $\phi$ meson has been measured by the NA11 Collaboration, at
the CERN-SPS, studying collisions of 120~GeV protons on beryllium and
tantalum targets, in the $\phi \to K^+K^-$ decay
channel~\cite{Bailey:1983uh}. They obtained a value $\alpha_\phi =
0.86 \pm 0.02$ (with a systematic uncertainty estimated to be around
three times smaller than the statistical one) for $\phi$ mesons
produced in the phase space window $0 < x_\mathrm{F} < 0.3$ and $\pt <
1$~GeV/$c$.
The BIS-2 Collaboration, at Serpukhov, measured the $\alpha$ value of
the $\phi$ mesons produced in collisions induced by neutrons of
energies between 30 and 70~GeV colliding on carbon, aluminum and
copper targets, considering the $\phi \to K^+ K^-$
decay channel~\cite{Aleev:1990tf}, obtaining $\alpha_\phi = 0.81 \pm
0.06$ for $\phi$ mesons produced with $x_\mathrm{F} > 0$ and $\pt <
1$~GeV/$c$ (without mentioning systematic errors).  More recently, as
already mentioned in the the analysis of the $\pt$ spectra, the HERA-B
experiment at DESY studied $\phi$-meson production in p-C, p-Ti and p-W
collisions at $\snn = 41.6$~GeV~\cite{Abt:2006wt}. HERA-B found a
value $\alpha_\phi = 0.96 \pm 0.02$ in the phase space covered by the
detector, $-0.7 < y_\mathrm{cms} < 0.25$ and $0.3 < \pt^2 <
12$~(GeV/$c$)$^2$. Extrapolating the measurements to zero $\pt$, the
value for $\alpha_\phi$ decreases to $\alpha_\phi = 0.91 \pm 0.02$
(where the error includes the systematic uncertainty). 
The HERA-B results also suggest an increase of the $\alpha_\phi$ parameter with $\pt$ 
(also reported for the $K^{*0}$ and for the $\bar{K}^{*0}$
mesons), with values of $\alpha_\phi$ as large
as 1.1 for $\pt^2 \sim 10$~(GeV/$c$)$^2$. A remarkable agreement is observed between the
$\pt$-dependence of $\alpha_\phi$ measured in the present analysis 
and the results of HERA-B, as shown in the right-top panel
of~\figurename~\ref{fig:alphaVsPt}. The KEK-PS E325
experiment~\cite{Tabaru:2006az} also measured the $\alpha$ parameter
for the $\phi$ meson, studying its $e^+e^-$ decay in p-C and p-Cu
collisions at $\snn = 5.1$~GeV. In the region of $0.9 < y < 1.7$
and $\pt < 0.75$~GeV/$c$ they found $\alpha_\phi = 0.937 \pm 0.049 \pm
0.018$, larger than the NA60 measurement in the same $\pt$ range,
but still compatible within the uncertainties. 
Summarising, despite the different collision systems, energies and
also kinematic regions, there is a substantial agreement between the $\alpha_\phi$ 
value extracted from the present analysis and the values from 
previous measurements.

~\\
\noindent A comparison of the results on the $\alpha_\eta$ parameter can be
established with the data published by the
CERES-TAPS Collaboration~\cite{Agakishiev:1998mw}: good
agreement is observed between the two sets of data points,
as shown in the right-bottom panel of~\figurename~\ref{fig:alphaVsPt}.
No similar comparison is possible for the $\omega$, due to the lack
of available measurements. The only available value for
the $\omega$ meson is reported by the KEK-PS E325 Collaboration, already
cited for the $\phi$ meson, which measured $\alpha_\omega = 0.710 \pm 0.021 \pm 0.037$, compatible
with the value of $\alpha$ for the total inelastic p-A cross section
and somehow smaller than our result.

\section{Particle cross-section ratios in full phase space}

\label{particleRatios}

\noindent We finally consider the nuclear dependence of the 
particle cross-section ratios in full phase space, assuming the $\omega$ meson as the reference.
The results are compiled in \tablename~\ref{tab:values_ratio}.

\begin{figure*}[hbt] 
   \begin{center} 
   \includegraphics[width=0.44\textwidth]{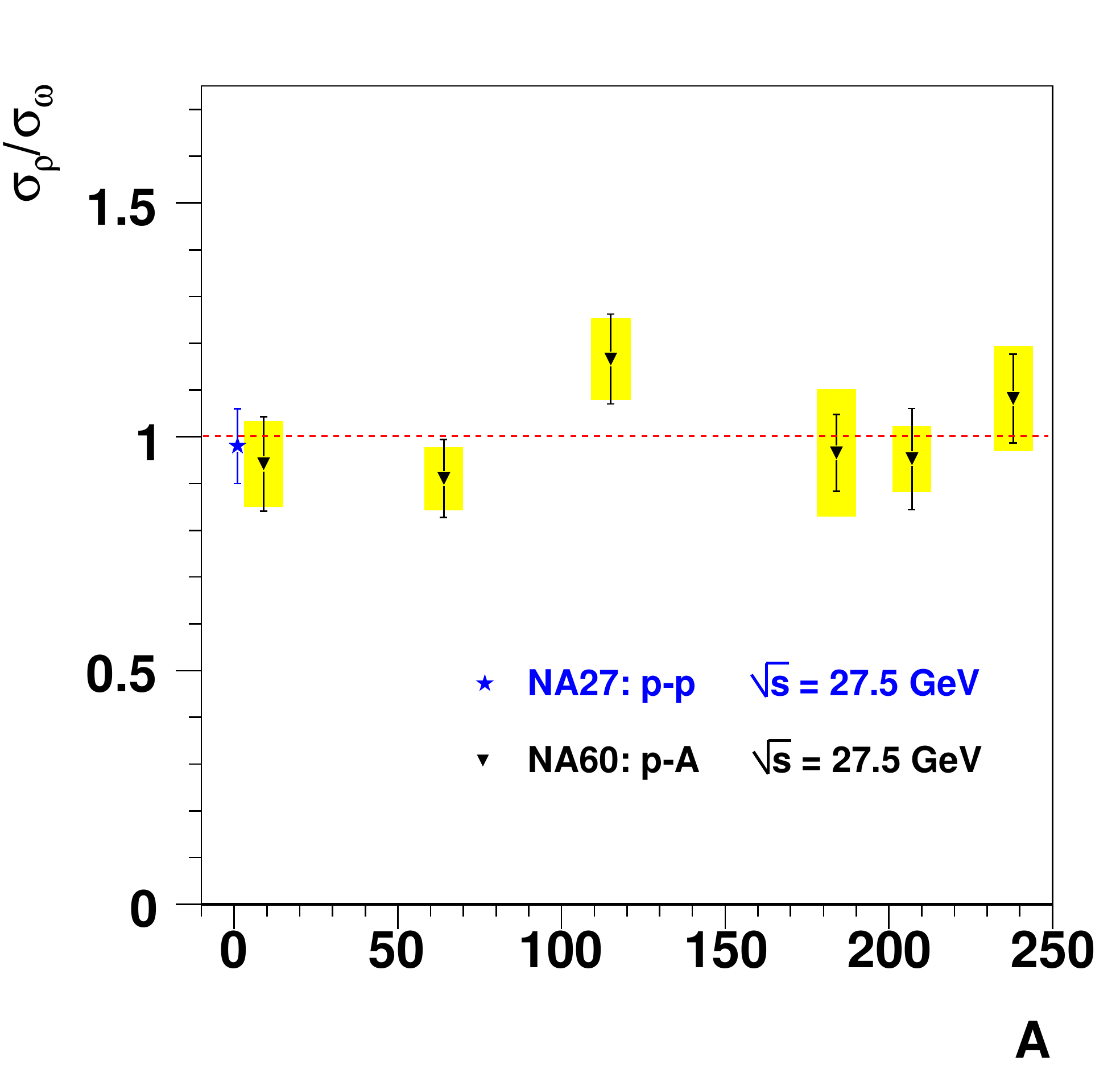} \hspace{0.04\textwidth} 
   \includegraphics[width=0.44\textwidth]{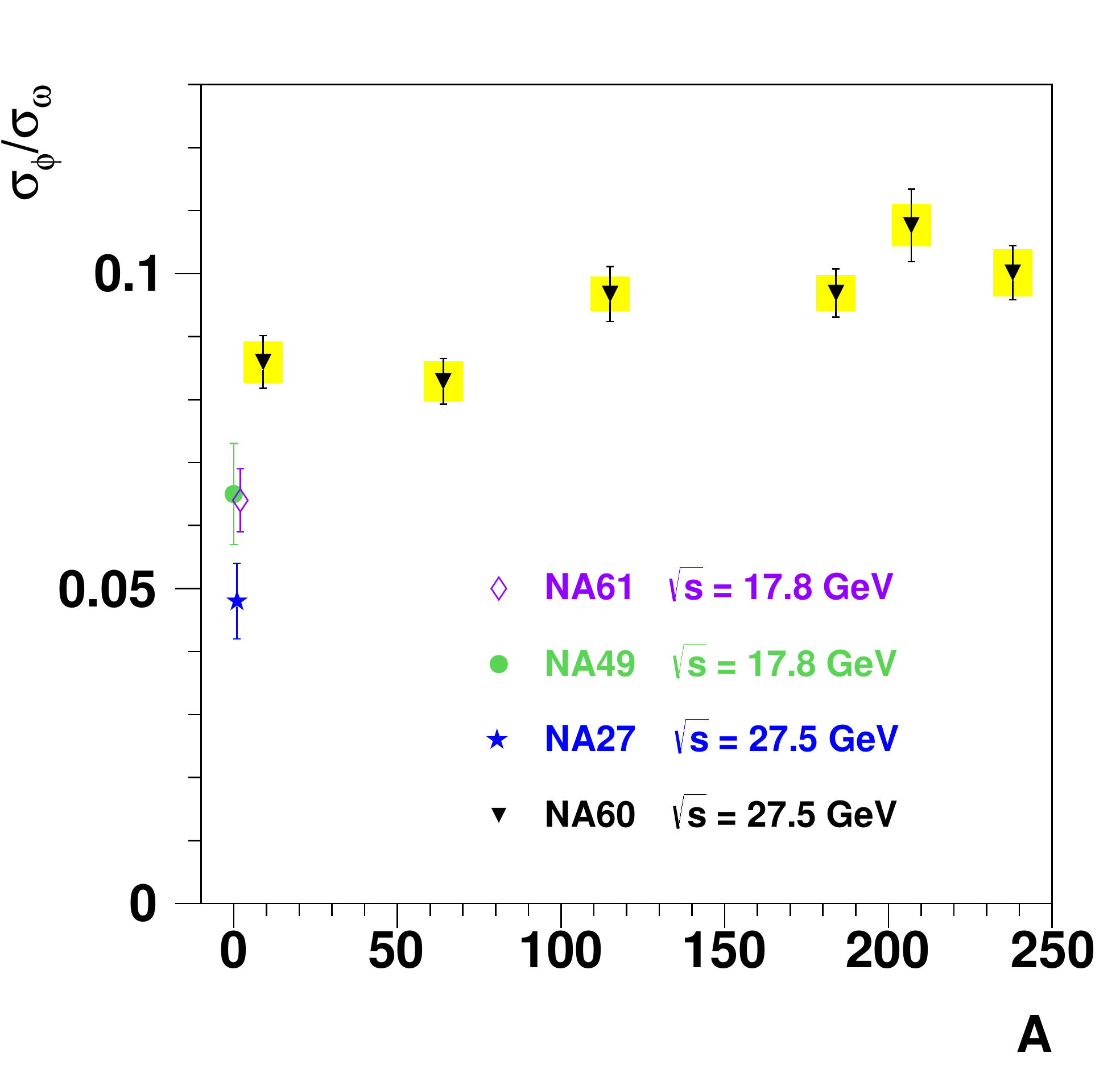} 
    \end{center} 
\caption[\textwidth]{Left: production cross section ratio
$\sigma_\rho/\sigma_\omega$ as a function of A. Right:  $\sigma_\phi/\sigma_\omega$  as a function of~A.
The ratios refer to
the full phase space, the error bars and shadowed boxes account for
statistical and systematic uncertainties, respectively.}
\label{fig:particleRatiosOmegaRhoPhi}
\end{figure*}

The $\sigma_\rho/\sigma_\omega$ ratio --- shown in the left panel of
\figurename~\ref{fig:particleRatiosOmegaRhoPhi} --- appears to be flat
with~A. The ratio averaged over
the different targets, indicated by a horizontal line, 
is $\sigma_\rho / \sigma_\omega = 1.00 \pm 0.04~\mathrm{(stat.)}
\pm 0.04~\mathrm{(syst.)}$, in agreement with the ratio
$\rho/\omega=0.98\pm0.08$ measured in p-p collisions at
$\sqrt{s}=27.5$~GeV by the NA27 experiment~\cite{AguilarBenitez:1991yy}, also shown in the figure. 

\noindent The $\sigma_\phi/\sigma_\omega$ ratio is shown in the right panel of
\figurename~\ref{fig:particleRatiosOmegaRhoPhi}. The trend of the NA60 data
points points to a $\sim 20\,\%$ increase of the $\sigma_\phi/\sigma_\omega$ ratio 
from Beryllium to Uranium, as to be expected from the difference between the $\pt$-integrated 
$\alpha_\omega$ and $\alpha_\phi$ coefficients, see section~\ref{nuclearDepPtIntegrated}.
The $\sigma_\phi/\sigma_\omega$ ratio in p-p collisions can be derived from the data
of the NA27~\cite{AguilarBenitez:1991yy}, NA49~\cite{Afanasev:2000uu,Alt:2005zq} and 
NA61/SHINE~\cite{NA49:2017mtg} experiments. The corresponding values
are also reported in \figurename~\ref{fig:particleRatiosOmegaRhoPhi}. 
Given the low atomic number of the Be nucleus (with p-Be often historically taken as
a proxy of p-p collisions) one might expect the $\sigma_\phi/\sigma_\omega$ ratio to
show similar values in p-Be and p-p: this scenario is, however, not favoured by the data presented 
in \figurename~\ref{fig:particleRatiosOmegaRhoPhi}, which suggest instead a discontinuity in the
evolution of the $\sigma_\phi/\sigma_\omega$ ratio from pp to p-A. The origin
of this behaviour, should it be attributed 
to a genuine physical mechanism, cannot be further investigated on the basis of the currently available data.

\begin{table*}[t]
  \vspace{0.2cm}
  \begin{center}

    \begin{tabular}{ c | c | c | c }

      \hline \hline

      \raisebox{-4pt}[0pt][10pt]{\textbf{A}} & 
      \raisebox{-4pt}[0pt][10pt]{$\boldsymbol{\sigma_\eta/\sigma_\omega}$} &
      \raisebox{-4pt}[0pt][10pt]{$\boldsymbol{\sigma_\rho/\sigma_\omega}$} &
      \raisebox{-4pt}[0pt][10pt]{$\boldsymbol{\sigma_\phi/\sigma_\omega}$} \\

      \hline \hline

      \raisebox{-4pt}[0pt][10pt]{Be} & 
      \raisebox{-4pt}[0pt][10pt]{~~~$1.42 \pm 0.08 \pm 0.17$~~~} &
      \raisebox{-4pt}[0pt][10pt]{~~~$0.94 \pm 0.10 \pm 0.09$~~~} &
      \raisebox{-4pt}[0pt][10pt]{~~~$0.086 \pm 0.004 \pm 0.003$~~~}\\

      \raisebox{-4pt}[0pt][10pt]{Cu} & 
      \raisebox{-4pt}[0pt][10pt]{~~~$1.51 \pm 0.07 \pm 0.16$~~~} &
      \raisebox{-4pt}[0pt][10pt]{~~~$0.91 \pm 0.08 \pm 0.07$~~~} &
      \raisebox{-4pt}[0pt][10pt]{~~~$0.083 \pm 0.004 \pm 0.003$~~~}\\

      \raisebox{-4pt}[0pt][10pt]{In} & 
      \raisebox{-4pt}[0pt][10pt]{~~~$1.89 \pm 0.09 \pm 0.21$~~~} &
      \raisebox{-4pt}[0pt][10pt]{~~~$1.17 \pm 0.10 \pm 0.09$~~~} &
      \raisebox{-4pt}[0pt][10pt]{~~~$0.097 \pm 0.004 \pm 0.003$~~~}\\

      \raisebox{-4pt}[0pt][10pt]{W} & 
      \raisebox{-4pt}[0pt][10pt]{~~~$1.87 \pm 0.08 \pm 0.25$~~~} &
      \raisebox{-4pt}[0pt][10pt]{~~~$0.97 \pm 0.08 \pm 0.14$~~~} &
      \raisebox{-4pt}[0pt][10pt]{~~~$0.097 \pm 0.004 \pm 0.003$~~~}\\

      \raisebox{-4pt}[0pt][10pt]{Pb} & 
      \raisebox{-4pt}[0pt][10pt]{~~~$2.34 \pm 0.12 \pm 0.20$~~~} &
      \raisebox{-4pt}[0pt][10pt]{~~~$0.95 \pm 0.02 \pm 0.11$~~~} &
      \raisebox{-4pt}[0pt][10pt]{~~~$0.108 \pm 0.006 \pm 0.003$~~~}\\

      \raisebox{-4pt}[0pt][10pt]{U} & 
      \raisebox{-4pt}[0pt][10pt]{~~~$2.04 \pm 0.09 \pm 0.27$~~~} &
      \raisebox{-4pt}[0pt][10pt]{~~~$1.08 \pm 0.09 \pm 0.11$~~~} &
      \raisebox{-4pt}[0pt][10pt]{~~~$0.100 \pm 0.004 \pm 0.004$~~~}\\

      \hline \hline

    \end{tabular}
  \end{center}
  \caption[\textwidth]{Results for the cross section ratios $\sigma_\eta/\sigma_\omega$, $\sigma_\rho/\sigma_\omega$ and $\sigma_\phi/\sigma_\omega$ as a function of the production target.}
  \label{tab:values_ratio}
\end{table*}

\begin{figure*}[hbt] 
   \begin{center} 
   \includegraphics[width=0.44\textwidth]{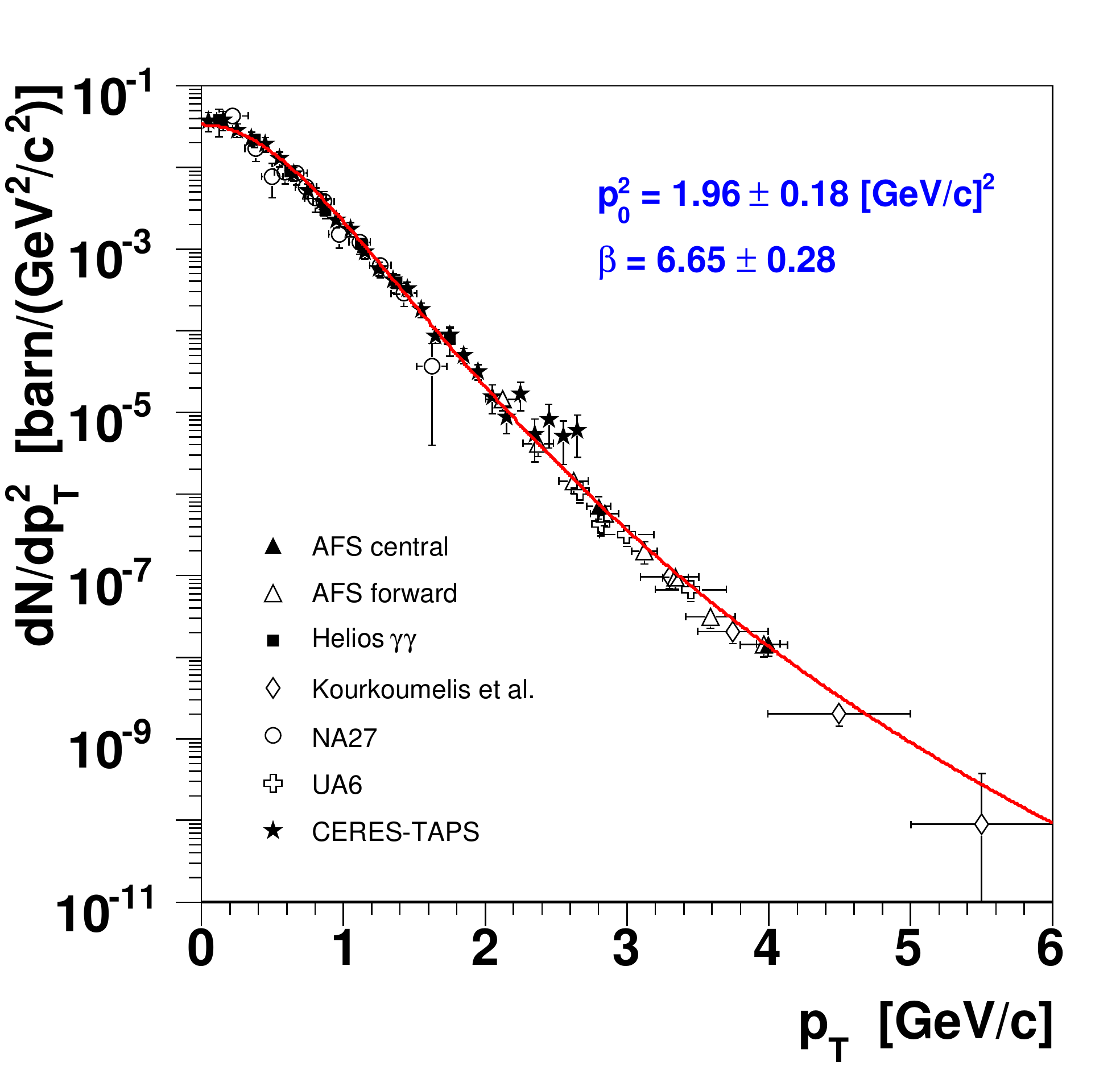} \hspace{0.04\textwidth}
   \includegraphics[width=0.44\textwidth]{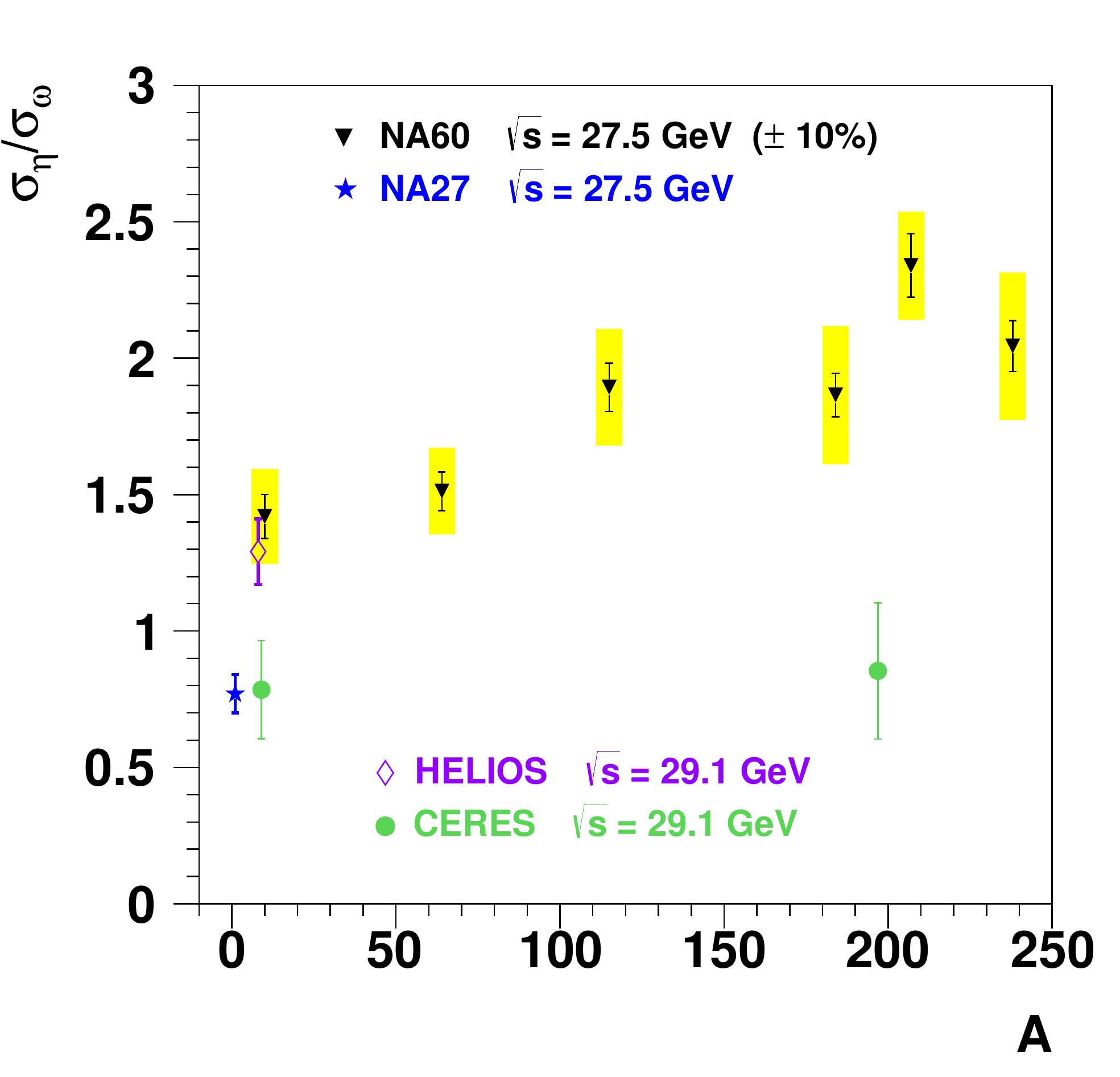}
    \end{center} 
\caption[\textwidth]{Left: compilation of the available $\pt$
measurements for the $\eta$ and fit with the power-law function~(\ref{eqn:pt2_function_2}). Right: cross section ratio
$\sigma_\eta/\sigma_\omega$ as a function of~A. The ratios refer to
the full phase space, the error bars and shadowed boxes account for
statistical and systematic uncertainties, respectively. An additional 10\,\% systematics uncertainty common to the NA60 points, coming from the uncertainty on the extrapolation factor to full phase space of the $\eta$ measurement, is represented by the band on the vertical axis. }
\label{fig:particleRatiosEta}
\end{figure*}

\noindent Special care must be taken when considering the $\sigma_\eta/\sigma_\omega$ ratio. Indeed, since the $\eta$
is measured in the present analysis via its Dalitz decay, and only for $\pt>0.6$~GeV/$c$ of the muon pair, 
an extrapolation of the $\eta$ $\pt$ spectrum down to zero~$\pt$ was needed, to recover the $\eta$ cross section in the full phase space. 
This full phase space extrapolation is based on a variety of available measurements, 
compiled in the plot shown on the left side of
\figurename~\ref{fig:particleRatiosEta} 
(see~\cite{Veenhof:1993xt,Agakishiev:1998mw,AguilarBenitez:1991yy,Kourkoumelis:1979ts,Akesson:1983xq,Akesson:1985za,Akesson:1986nx,Antille:1987kr}). 
On these data, a fit is performed with the power-law function given by Eq.~(\ref{eqn:pt2_function_2}), providing
an excellent description of the points in the whole $\pt$ range. 
The $\sigma_\eta/\sigma_\omega$ ratio extrapolated to the full phase space is shown in the right panel of \figurename~\ref{fig:particleRatiosEta}. 
The systematic uncertainty from the extrapolation of the $\eta$ measurement to full phase space amounts to $\sim$10\,\%, point-to-point fully correlated. 
It should be remarked that the considered extrapolation factor does not depend on the nuclear target, and it is mainly based on measurements 
in elementary collisions (pp, $\bar{\mathrm{p}}$p or p-Be). As a consequence, the NA60 results for the $\pt$-integrated $\sigma_\eta/\sigma_\omega$ ratio
are properly normalised at the p-Be point, while the trend versus~A should be ascribed to the nuclear dependence of the ratio 
in the range $\pt > 0.6$~GeV/$c$, where particles are measured, as to be expected from the difference between the $\alpha_\omega$ and $\alpha_\eta$ 
coefficients in this same kinematic range, see section~\ref{nuclearDepPtIntegrated}.

\noindent The $\sigma_\eta/\sigma_\omega$ cross-section ratio was also measured by NA27 in p-p
collisions at $\sqrt{s}=27.5$~GeV~\cite{AguilarBenitez:1991yy}, by HELIOS in
p-Be collisions at $\snn=29.1$~GeV~\cite{Akesson:1994mb} and by
CERES-TAPS in p-Be and p-Au collisions at
$\snn=29.1$~GeV~\cite{Agakishiev:1998mw}. The corresponding
measurements are shown in the right panel of
\figurename~\ref{fig:particleRatiosEta} together with the NA60 results. 
The measurement of $\sigma_\eta/\sigma_\omega$ by the HELIOS experiment was derived
from the $\sigma_\eta/(\sigma_\rho+\sigma_\omega)$ ratio measured in p-Be collisions
through the detection of both dielectrons and dimuons, exploiting the
capability to fully reconstruct the Dalitz decays. Neglecting possible
$\rho/\omega$ interference effects, HELIOS obtained
$\sigma_\eta/(\sigma_\rho+\sigma_\omega) = 0.54 \pm 0.05$ from the
$e^+e^-$ data and $0.52 \pm 0.06$ from the $\mu^+\mu^-$ data. These
results were based on a $(1+\cos^2 \theta)$ decay angle distribution
to extrapolate the $\rho$ and $\omega$ measurements to full phase
space. If \mbox{HELIOS} had assumed a uniform dimuon decay angle distribution
for the $\rho$ and $\omega$ mesons, as done in the present analysis,
the full phase space $\rho$ and $\omega$ cross sections would have
been $\sim20\,\%$ larger, resulting in a $\sigma_\eta/\sigma_\omega$
ratio around $1.28 \pm 0.12$, in  agreement with the NA60
value. Finally, the CERES-TAPS $\sigma_\eta/\sigma_\omega$ measurements in p-Be and p-Au are based on
the $\eta\to\gamma\gamma$ and $\omega\to\pi^0\gamma$ decay channels for $\pt > 0$, resulting in
the two points also shown in the right panel of
\figurename~\ref{fig:particleRatiosEta}. 

\noindent The comparison between the p-A results from NA60 and the p-p point from NA27 suggests
a discontinuity in the evolution of the $\sigma_\eta/\sigma_\omega$ ratio from p-p to p-A,
similar to what remarked for the $\sigma_\phi/\sigma_\omega$ ratio. The HELIOS measurement
in p-Be collisions agrees with the NA60 data, while the p-Be measurement of CERES-TAPS looks
compatible with the p-p point from NA27. At large~A, the NA60 data cannot be directly compared to the CERES-TAPS 
point in p-Au, because of the different range of $\pt$ covered by the two measurements in evaluating
the nuclear dependence of the $\eta$ cross section.
The CERES-TAPS data, taken alone, are not conclusive about the nuclear dependence 
of the $\sigma_\eta/\sigma_\omega$ ratio, because of the only two nuclear targets
available and the rather large statistical uncertainties.

\section{Conclusions}

\noindent In this letter we reported on a comprehensive study of low-mass dimuon production in p-A interactions at 400~GeV, ranging from 
p-Be to p-U, measured with the NA60 apparatus at the CERN SPS. 
The analysis of the $\pt$ spectra for the $\rho/\omega$ and $\phi$ mesons has shown that the observed distributions cannot be explained by a 
thermal-like exponential function in the full $\pt$ range accessed, while a good description of the whole observed spectrum is provided by the power-law 
parametrisation $\d N/\d \pt^2 \propto \left( 1 + \pt^2/p_0^2 \right)^{-\beta}$. 
The nuclear dependence of the cross sections of the $\eta$, $\omega$ and $\phi$ mesons has been found to be compatible with the power law 
$\sigma_\mathrm{pA} \propto \mathrm{A}^\alpha$, with a clear rising trend of the $\alpha$ parameters observed as a function of~$\pt$.
The observed, approximate hierarchy $\alpha_\eta \approx \alpha_\phi > \alpha_\omega$ could be attributed to the $s\bar{s}$ component
in the quark wave function of the $\phi$ and $\eta$ mesons, resulting in a harder production mechanism with respect to the $\omega$ meson.
The measurement of the nuclear dependence of the production cross section ratios points to a moderate 
rising trend as a function of~A of both the $\sigma_\eta/\sigma_\omega$ and $\sigma_\phi/\sigma_\omega$ --- 
reflecting the observed difference in the $\alpha$ parameters of the $\eta$, $\omega$ and $\phi$ mesons ---
while the $\sigma_\rho/\sigma_\omega$ ratio has been found to be 
almost independent of the production target, and compatible with unity, in agreement with the available measurement in p-p. 

\bibliographystyle{unsrt}
\bibliography{na60_pA2004_nuclear-dependence}

\end{document}